\documentclass[a4paper,fleqn,usenatbib]{mnras}

\usepackage{newtxtext,newtxmath}

\usepackage[T1]{fontenc}
\usepackage{ae,aecompl}

\usepackage{amsmath}	
\usepackage{amssymb}	
\usepackage{graphicx,color,afterpage}
\usepackage{times,subcaption,rotating,tikz}
\usepackage{float}

\def\kms{km\,s$^{-1}$}

\def\msun{M$_{\odot}$}

\def\arcsec{$^{\prime \prime}$}
\definecolor{Mygrey}{gray}{0.75}

\newcommand{\ltsimeq}{\raisebox{-0.6ex}{$\,\stackrel{\raisebox{-.2ex}{$\textstyle <$}}{\sim}\,$}}

\newcommand{\farc}{\mbox{\ensuremath{.\!\!^{\prime\prime}}}}

\mathchardef\mhyphen="2D

\usepackage[compact]{titlesec}
\titlespacing{\section}{0pt}{*2}{*1}

\title[Gas properties of simulated PSBs]{Evolution of the cold gas properties of simulated post-starburst galaxies} 
\author[Timothy A. Davis et al.]{\parbox{\textwidth}{Timothy A. Davis,$^{1}$\thanks{E-mail: \texttt{DavisT@cardiff.ac.uk}} Freeke van de Voort,$^{2,3}$ Kate Rowlands,$^{4}$ Stuart McAlpine,$^{5,6}$\\ Vivienne Wild$^{7}$ and Robert A. Crain$^{8}$
}
\vspace{0.4cm}\\
\parbox{\textwidth}{$^{1}$School of Physics \&\ Astronomy, Cardiff University, Queens Buildings, The Parade, Cardiff, CF24 3AA, UK\\
$^{2}$Heidelberg Institute for Theoretical Studies, Schloss-Wolfsbrunnenweg 35, D-69118 Heidelberg, Germany\\
$^{3}$Astronomy Department, Yale University, PO Box 208101, New Haven, CT 06520-8101, USA\\
$^{4}$Department of Physics \&\ Astronomy, Johns Hopkins University, Bloomberg Centre, 3400 N. Charles St., Baltimore, MD 21218, USA\\
$^{5}$Institute for Computational Cosmology, Department of Physics, Durham University, South Road, Durham DH1 3LE, UK\\
$^{6}$Department of Physics, University of Helsinki, Gustaf H\"allstr\"omin katu 2a, 00560 Helsinki, Finland\\
$^{7}$School of Physics \&\ Astronomy, University of St Andrews, North Haugh, St Andrews, Fife, KY16 9SS, UK\\
$^{8}$Astrophysics Research Institute, Liverpool John Moores University, 146 Brownlow Hill, Liverpool L3 5RF, UK
}}
\begin{document}
\date{Accepted 2019 January 14. Received 2019 January 14; in original form 2018 September 26.}
\pagerange{\pageref{firstpage}--\pageref{lastpage}} \pubyear{2018}
\maketitle
\label{firstpage}
\begin{abstract}
Post-starburst galaxies are typically considered to be a transition population, en route to the red sequence after a recent quenching event. Despite this, recent observations have shown that these objects typically have large reservoirs of cold molecular gas. In this paper we study the star-forming gas properties of a large sample of post-starburst galaxies selected from the cosmological, hydrodynamical EAGLE simulations. These objects resemble observed high-mass post-starburst galaxies both spectroscopically and in terms of their space density, stellar mass distribution and sizes. We find that the vast majority of simulated post-starburst galaxies have significant gas reservoirs, with star-forming gas masses $\approx10^9$ \msun, in good agreement with those seen in observational samples. The simulation reproduces the observed time evolution of the gas fraction of the post-starburst galaxy population, with the average galaxy losing $\approx$90 per cent of its star-forming interstellar medium in only $\approx$600 Myr. A variety of gas consumption/loss processes are responsible for this rapid evolution, including mergers and environmental effects, while active galactic nuclei play only a secondary role. The fast evolution in the gas fraction of post-starburst galaxies is accompanied by a clear decrease in the efficiency of star formation, due to a decrease in the dense gas fraction. We predict that forthcoming ALMA observations of the gas reservoirs of low-redshift post-starburst galaxies will show that the molecular gas is typically compact and has disturbed kinematics, reflecting the disruptive nature of many of the evolutionary pathways that build up the post-starburst galaxy population. 
\end{abstract}


\section{\uppercase{Introduction}}
\noindent Ever since the first large-scale surveys revealed the diversity present in the galaxy population, understanding morphology has formed a key part of the study of galaxy evolution. Blue, star-forming galaxies appear very common in the early universe, while more quiescent early-type galaxies (with a lenticular or elliptical morphology) begin to dominate at more recent times \citep[e.g.][]{2007ApJ...665..265F}. By redshift zero early-type galaxies contain $\approx$70 per cent of the stellar mass \citep[][]{2014MNRAS.444.1647K}, and dominate the galaxy population in dense environments \citep[e.g.][]{1974ApJ...194....1O,1980ApJ...236..351D}. Given the strong time evolution of these two populations a key aspect of modern astrophysics involves identifying and quantifying the different mechanisms that quench and morphologically transform galaxies.

As with many questions of this type, identifying the population of objects in transition from one type to another is crucial. 
One such candidate transition population are the post-starburst (PSB) galaxies, which are thought to have been caught in the midst of a rapid transition from star-forming
to quiescent. 
These objects have very low star formation rates (as indicated by their lack of significant nebular emission lines), but strong Balmer absorption lines
(indicative of a large population of A- and F-type stars).  This suggests that these galaxies have only ceased forming stars in the past billion years, potentially after a large burst of star formation \citep{1983ApJ...270....7D,1987MNRAS.229..423C}.
PSBs are thus among the best candidates for objects that have been quenched rapidly in their recent past.  Some studies suggest around half of all galaxies have experienced quenching in this manner on their way to the red sequence \citep{2007ApJS..173..342M,2009MNRAS.395..144W,2015MNRAS.450..435S,2016MNRAS.463..832W}. Understanding which processes act to create such objects is thus clearly vital. 

In order to quench a star-forming galaxy, and turn it into a quiescent system, a large part of the cold atomic and molecular gas reservoir needs to be rendered inhospitable to star formation, either through gas heating or removal.
Until recently, however, the cold gas reservoirs of PSBs had not been systematically surveyed. The studies of \cite{2015ApJ...801....1F} and \cite{2015MNRAS.448..258R} were among the first to show that significant reservoirs of molecular
gas remain in PSBs. Further studies followed, shedding light on the molecular \citep[e.g.][]{2016ApJ...827..106A,2017MNRAS.469.3015Y,2017ApJ...846L..14S}, atomic \citep{2001AJ....121.1965C,2013MNRAS.432..492Z,2017A&A...600A..80K,2018MNRAS.478.3447E}, and dust \citep{2018ApJ...855...51S} properties of such systems. Taken together, these studies show that large interstellar medium (ISM) masses are very common in the PSB galaxy population.  

While studies of molecular gas reservoirs in early-type galaxies \cite[e.g.][]{2007MNRAS.377.1795C,2011MNRAS.414..940Y,2016MNRAS.455..214D} have shown that there is no need to remove the entire molecular reservoir of a galaxy during  quenching, the large molecular gas reservoirs present in PSB galaxies ($10^9$--$10^{10}$ \msun) came as a surprise. 
Overall, it seems that star formation quenching in PSBs is not a result of gas exhaustion; there is no need to remove the molecular reservoir of a galaxy to trigger a PSB episode that could lead to quenching and galaxy transition.

How a galaxy, which still has a significant cold gas reservoir, can become quiescent is an open question, which we aim to address here by studying the star-forming gas of PSB galaxies within the cosmological, hydrodynamic EAGLE simulation \citep{Schaye2015, Crain2015}. 
Pawlik et al.\ submitted recently studied a small sample of PSBs selected from EAGLE, and showed clearly that these objects follow a diverse set of evolutionary pathways. They did not, however, consider the gas properties of these systems in detail. Here we concentrate on understanding the cold gas in PSBs within EAGLE, selecting a large sample of objects at a range of redshifts. 
 By comparing the observed PSB galaxies with those selected in the simulation we aim to shed light on their properties in the full cosmological context, and make predictions for resolved observations of the PSB galaxy population with e.g.\ the Atacama Large Millimeter/submillimeter Array (ALMA) and optical integral field studies such as MaNGA \citep{2015ApJ...798....7B}.

In Section \ref{data} we describe the pertinent details of the EAGLE simulation we use, the selection of PSBs, and the observed comparison datasets. In Section \ref{results} we describe the main results, before discussing and concluding in Section \ref{conclude}. Throughout the paper we use a $\Lambda$CDM cosmology with parameters $\Omega_\mathrm{m} = 1 - \Omega_\Lambda = 0.307$, $\Omega_\mathrm{b} = 0.04825$, $h = 0.6777$, $\sigma_8 = 0.8288$, $n = 0.9611$ \citep{Planck2014}. 

\section{\uppercase{Method}}
\label{data}

\begin{figure*}
        \centering
  \includegraphics[height=6cm,angle=0,clip,trim=0cm 0cm 0cm 0.0cm]{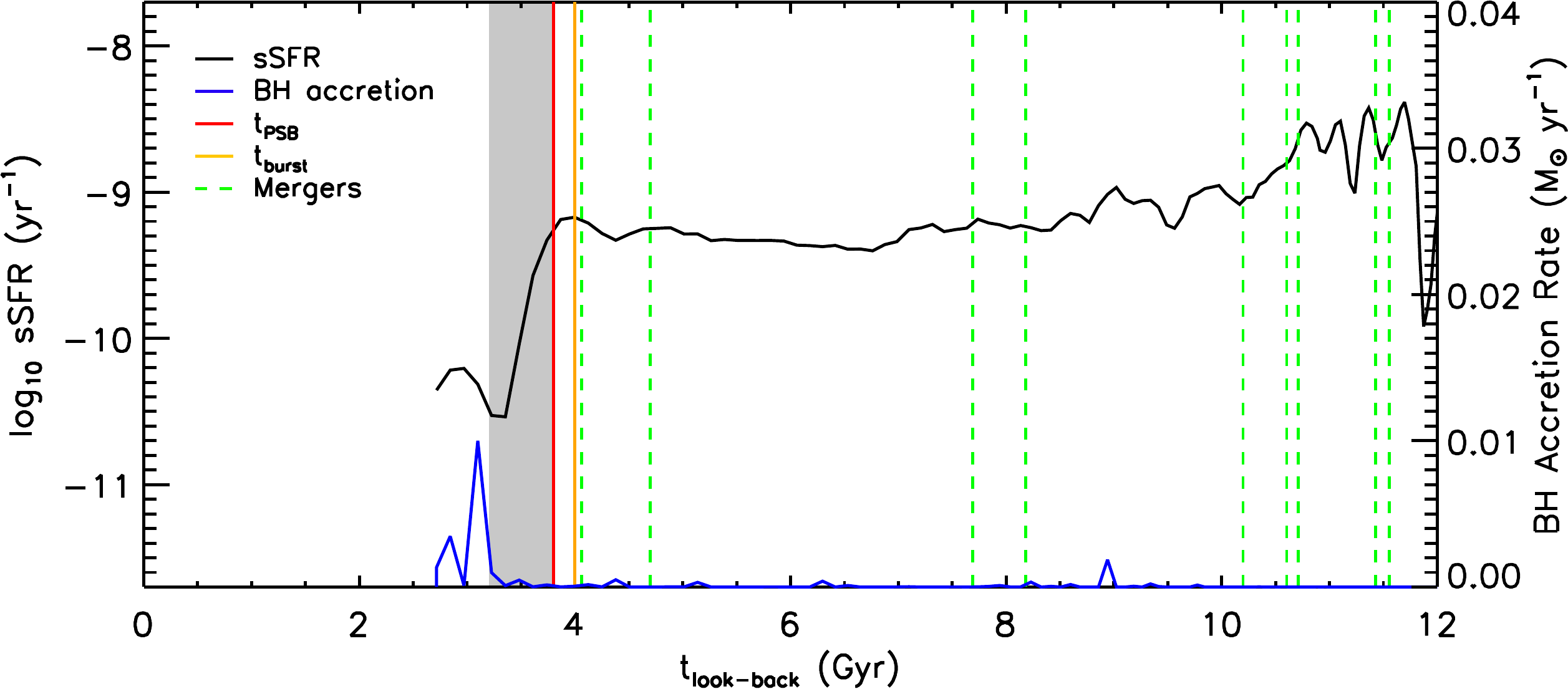}
        \caption{Specific star formation history (black line) for an example post-starburst galaxy. The red vertical line indicates $t_{\rm PSB}$, determined as described in Section \ref{howweselect}, while the grey shaded area indicates the PSB period ($t_{\rm PSB}$ to $t_{\rm PSB}-600$ Myr). $t_{\rm burst}$ is indicated by the orange line. Green vertical dashed lines show the coalescence time for each of the $>1$ per cent stellar mass fraction mergers this object experiences. Also shown as a blue curve is the instantaneous AGN accretion rate in this object. Our selection method cleanly identifies the large sSFR drop in this object, flagging it as a PSB candidate. }
        \label{ssfr_example}
 \end{figure*}%
\subsection{The EAGLE simulation suite}
\label{eagle_explain}

In this work we compare the observational properties of PSBs with simulated galaxies from the `Evolution and Assembly of GaLaxies and their Environments' (EAGLE) project \citep{Schaye2015, Crain2015, 2016A&C....15...72M}. This project consists of a large number of cosmological hydrodynamical simulations with different subgrid physics, simulation volumes, and resolution, which were run using a modified version of the smoothed particle hydrodynamics (SPH) code \textsc{gadget-3} \citep[last described in][]{Springel2005}. Here we make use of the high time resolution outputs (referred to as `snipshots') produced from the largest (100 co-moving Mpc on a side) cubic periodic volume simulation (\emph{``Ref-L100N1504''}), which includes both stellar and active galactic nucleus (AGN) feedback. A comprehensive description of this simulation, and of the database of galaxy properties extracted from it, can be found in \citet{Schaye2015,Crain2015, 2016A&C....15...72M}.

Due to numerical constraints the multi-phase structure of the ISM in EAGLE is not resolved. In this work, where we wish to compare the simulated galaxies' molecular ISM to observed samples, we assume that all of the \textit{star-forming} gas particles are molecular gas dominated. This has been shown to provide a reasonable estimate of the molecular gas mass \citep{Lagos2015}. In EAGLE star formation proceeds in `dense' gas ($\geq0.1$~cm$^{-3}$ at solar metallicity), and is modelled as in \citet{Schaye2008}. 
The star formation rate (SFR) per unit mass depends on the gas pressure, and is calibrated to reproduce the observed Kennicutt-Schmidt law \citep{Kennicutt1998}. 

Galaxies were found within the simulation using a Friends-of-Friends (FoF) algorithm to identify dark matter haloes, and the \textsc{subfind} \citep{Springel2001, Dolag2009} algorithm to find the gravitationally bound particles and to identify sub-haloes.  The sub-halo containing the minimum of the gravitational potential is referred to as the central, with the remainder referred to as satellites. Note that when the central and satellite are of a very similar mass there is ambiguity in the interpretation of this label. The centre of each sub-halo is defined by the position of its most bound particle. 
Stellar mass ($M_\mathrm{*}$) is measured using the star particles associated with the galaxy's sub-halo, but only including those within 30~proper kpc of the galaxy's centre in order to exclude intra-halo stars.  For each galaxy the star-forming gas mass is calculated by summing the masses of all star-forming gas particles, again within 30~kpc of the centre of each sub-halo. 

The sub-grid model parameters within these simulations were calibrated such that the output reproduces several key stellar properties of observed galaxies (e.g. the stellar mass function, and the distribution of stellar sizes at $z=0.1$; \citealt{Furlong2015,Furlong2017}). However, they were not tuned to reproduce any observable \textit{gas} properties of these systems. Nevertheless the reference simulations reproduce, for instance, the H\,\textsc{i} and H$_2$ masses and mass-fractions in high mass ($>10^{10}$~M$_\odot$) galaxies \citep{Lagos2015, Bahe2016, Crain2017}, and the distribution of H\,\textsc{i} masses with environment \citep{Marasco2016}. 
We note, however, that there may not be sufficient cool gas in low-mass galaxies \citep{Crain2017}. Due to this issue we concentrate on the properties of reasonably massive (M$_*>10^{9.5}$ \msun) PSBs in this work. 
    
\begin{figure*}
    \begin{subfigure}[t]{0.5\textwidth}
        \centering
  \includegraphics[height=6cm,angle=0,clip,trim=0cm 0cm 0cm 0.0cm]{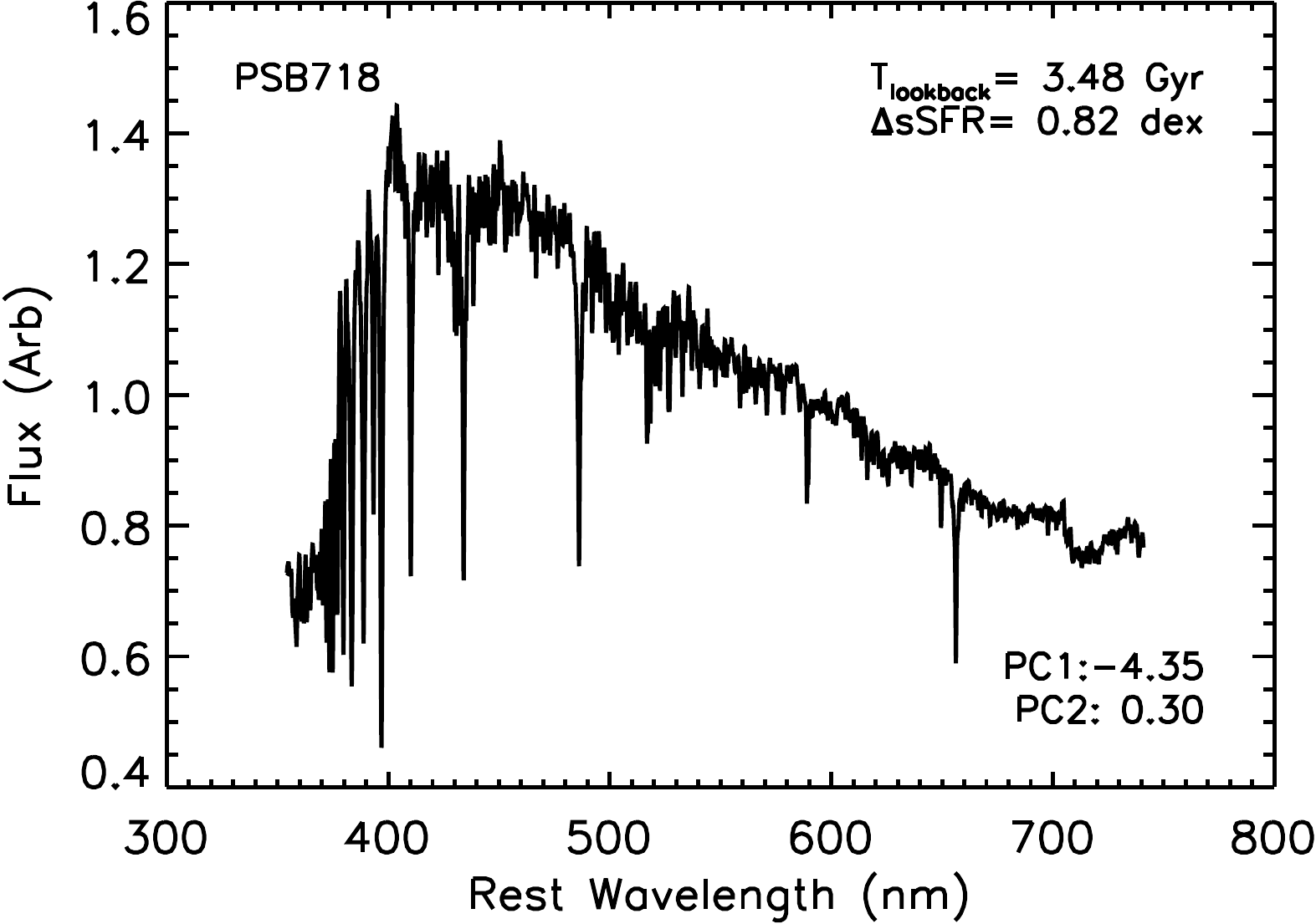}
        \caption{Mock optical spectrum of PSB718}
    \end{subfigure}%
    ~ 
    \begin{subfigure}[t]{0.5\textwidth}
        \centering
\includegraphics[height=6cm,angle=0,clip,trim=0cm 0cm 0cm 0.0cm]{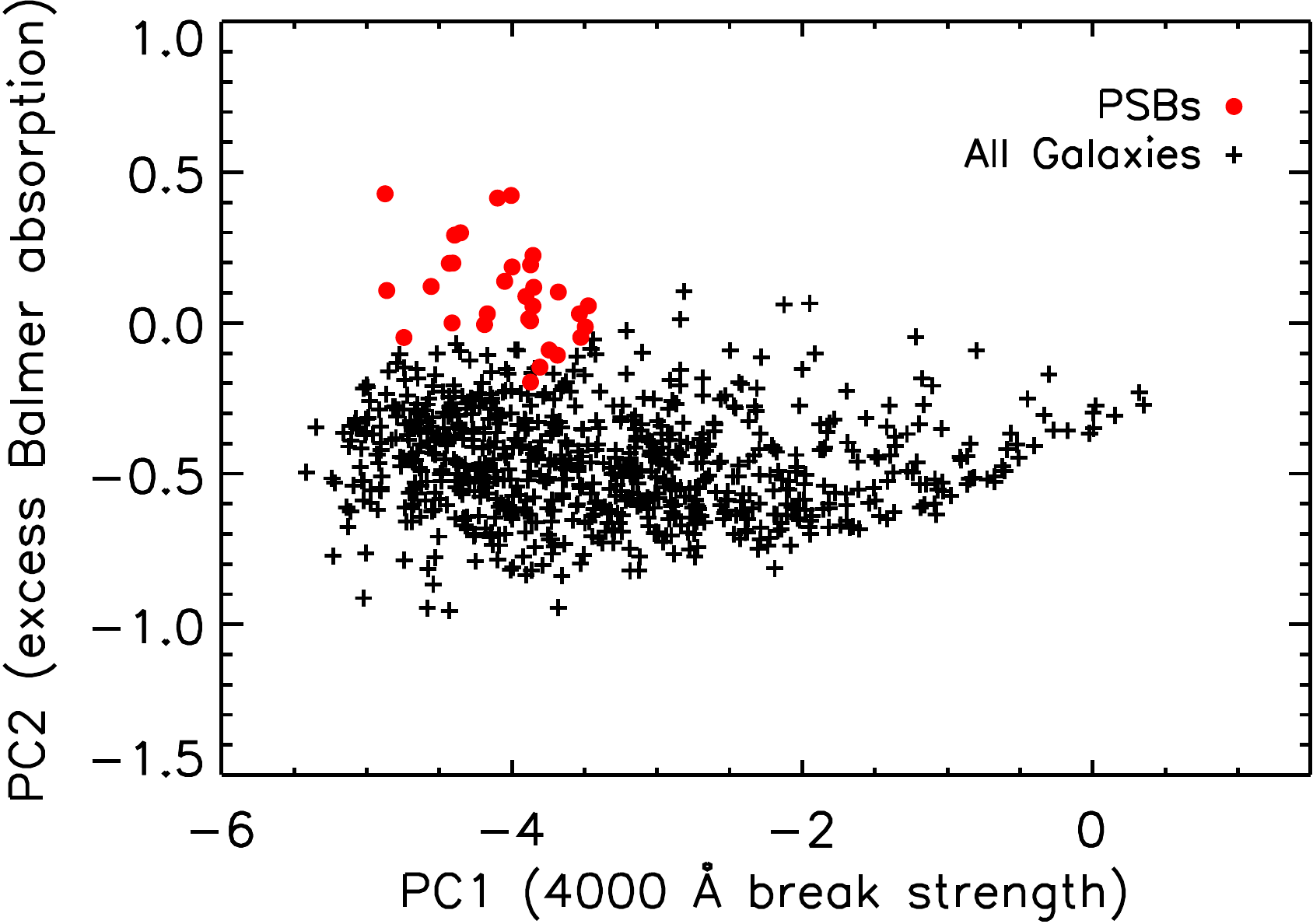}
        \caption{PC space for PSB and control galaxies at t$_{\rm look-back}$=3 - 4 Gyr.}
    \end{subfigure}
    \caption{ \textit{Left:} Mock spectrum of the example candidate galaxy from Figure \ref{ssfr_example}, produced 200 Myr after the onset of its star formation burst, clearly showing the strong Balmer absorption features typical of PSBs. \textit{Right:} Principle components for all of the simulated PSBs at t$_{\rm look-back}$=3 - 4 Gyr (red points; including PSB718) and the rest of the galaxy population in the simulation at that $z$, showing that the simulated PSB's clearly stand out in this space, and the vast majority would be selected by spectroscopic selection criteria.}
    \label{principle_comps}
\end{figure*}

\subsection{{Selecting PSBs}}
\label{howweselect}
Post-starburst galaxies have, by definition, undergone a strong recent truncation in their star formation history. Such truncations must be rapid in order to produce the observed spectral signatures, which (dependent on the observational method) can remain observable for $\approx$0.6 -- 1 Gyr \citep[e.g.][]{2010MNRAS.405..933W,2018arXiv180603301F}. 

In this work we select our PSB candidates by examining the star formation histories (derived from the formation times and initial masses of the stars present in each galaxy) for all massive (M$_*>10^{9.5}$ \msun) galaxies in the ``Ref-L100N1504'' simulation. In order to be selected as a PSB candidate, a galaxy must initially be star-forming, with a specific star formation rate (defined as the SFR per unit stellar mass; SFR/M$_*$) of a least $5\times10^{-11}$\,yr$^{-1}$, and then experience a sSFR drop of a least a factor of 5 within a period of 600\,Myr. We chose this timescale as it is well resolved by the eagle snipshots (which have a mean spacing of $\approx$60 Myr), and similar to the observability timescale for PSB features in optical spectra. {In order to check that our results remain robust to this choice, we varied both the sSFR drop ($>$5, 10 or 20) and period considered (400, 600 or 1000 Myr).  Although the number of objects selected changes in each case, the conclusions derived in this work remain robust.}
 Figure \ref{ssfr_example} shows an example star formation history for one simulated galaxy selected as a PSB by our method. 

We repeat this selection every 0.5 Gyr between $z=0$ and $z=2$, adding objects that meet our selection criteria (but were not selected in previous time steps) to our sample. This procedure crucially allows us to select high redshift objects that undergo mergers with larger objects at low redshifts, obscuring the drops in their combined $z=0$ star formation histories. We do not attempt to select objects beyond $z=2.5$ in this study. This selection results in a sample of 1244 post-starburst candidates. A small fraction of these objects ($\approx$2.7 per cent) have multiple PSB episodes (separated by more than 600 Myr) during their evolution.

For each object, we define $t_{\rm PSB}$ as the look-back time at the start of the 600 Myr window which contains the largest sSFR drop. We note, however, that this time is not defined in the same way as one might do observationally, based on age dating of starburst signatures. By inspection of the star formation histories of our objects, we find a typical {(population average)} delay of 200 Myr between the peak of a star formation burst and the start of the period of maximum sSFR suppression (see Section \ref{sec:sfe_evol}). Thus in order to provide a better observational comparison we define the look-back time to the starburst as $t_{\rm burst}\,=\,t_{\rm PSB}\,+\,200$ Myr.

\subsubsection{Defining a control sample}

As we are able to select a large sample of PSB candidates over a wide range of redshifts, it is important to construct a dedicated control sample with which to compare their properties. 
We therefore randomly selected 10 comparison objects per PSB, which have the same stellar and star-forming gas mass (within a tolerance of 0.1 dex) as the PSB candidate in the snipshot closest to $t_{\rm PSB}$. 

\subsection{Observational comparisons}
\label{obs_comp}
Many studies of PSBs have been conducted, using different selection methods, and targeting galaxies at a large range of redshifts. 
In this work we are primarily interested in understanding the evolution of the cold ISM in PSBs, however, and so the majority of our comparisons will be to the observational samples who have obtained molecular gas observations and where burst age estimates are available in the literature.    

The first sample we consider, from \citet{2015ApJ...801....1F}, uses a spectroscopic selection method, requiring objects to have strong stellar Balmer absorption lines (characteristic of a recent starburst), and little nebular emission (H$\alpha$ equivalent widths $<$ 3\,\AA\ in the rest frame).  The second sample, by \citet{2015MNRAS.448..258R}, uses the principle component analysis (PCA) based spectroscopic selection method  introduced by \citet{2007MNRAS.381..543W}, which selects high burst mass fraction, long duration starbursts/post-starbursts \citep{2018arXiv180603301F}. The third sample is the shocked post-starburst galaxy sample of \citet{2016ApJS..224...38A} (with molecular gas measurements presented in \citealt{2016ApJ...827..106A}), who select galaxies with deep Balmer absorption lines and emission line ratios which are inconsistent with star formation (and which could be from shocks). We note that we do not compare to the sample of green-valley PSB galaxies which also host Seyfert-type AGN activity from \cite{2017MNRAS.469.3015Y}, because although molecular gas observations are available, age estimates for the PSB activity are not.

These data form the bulk of our comparison set, together comprising 74 galaxies between $z=0$ and $z=0.2$ with molecular gas observations, spanning two orders of magnitude in stellar mass (10$^{9.5}$ -- 10$^{11.5}$ \msun). Where we compare with other surveys beyond these (for instance in order to extend to higher redshifts) this is discussed explicitly in the text.

\section{\uppercase{Results}}
\label{results}

\subsection{Do EAGLE PSBs resemble those observed?}
\label{psb_comp}

We identified a large population of galaxies that recently had a large drop in sSFR within the simulation. However, observational studies of PSBs are not able to select their candidates using the star formation histories (SFHs) directly, and instead rely on observational proxies. Typically these selection methods revolve around obtaining galaxy spectra, and searching for strong A-type star signatures, such as a strong 4000\,\AA\ break, and strong Balmer absorption lines \citep{1983ApJ...270....7D}. 

In order to determine if our objects would be selected by such methods we created mock spectra of our simulated PSBs. 
Each stellar particle in our simulated galaxies has an associated age and metallicity. We used these parameters along with the (e-)MILES \citep{2010MNRAS.404.1639V,2016MNRAS.463.3409V} single-stellar population (SSP) library (with a Chabrier IMF and stellar isochrones from \citealt{2004ApJ...612..168P}) to generate a stellar emission spectrum for each particle, which we then blue/red-shifted to take into account the mean motion of the star particle around the galaxy centre of mass. We neglected the effects of $\alpha$-enhancement and dust extinction here, but do not expect this to substantially alter our results. The final spectrum of each galaxy was then derived as the mass-weighted sum of the individual particle spectra. A representative example is shown in the left panel of Figure \ref{principle_comps}.  This object underwent a 0.82~dex drop in its sSFR at $z\approx$\,0.28, and this spectrum (produced 200 Myr after the onset of this drop) shows clear PSB signatures (i.e. a very strong Balmer break, and excess strong Balmer absorption).

In order to quantify this further we calculated the spectral indices of \cite{2007MNRAS.381..543W} for our mock spectra. These indices are defined from a PCA performed on a set of \citet{2003MNRAS.344.1000B} model spectra with exponentially declining SFHs and additional superimposed random bursts. The PCA results in a set of eigenvectors ordered by the amount of variance that they account for in the library of spectra. The projection of any spectrum onto the eigenvectors results in a set of principal component amplitudes, which describe the strength of the features in the spectrum.  The first principal component amplitude (PC1) is related to the strength of the 4000 Å break (equivalent to the $D_{n}4000$ index), and PC2 is the excess Balmer absorption (of all Balmer lines simultaneously) over that expected based on the 4000\,\AA\ break strength. The advantage of this method is the significant improvement in signal-to-noise ratio over the traditional method of measuring the H$\delta$ absorption line alone. See \cite{2007MNRAS.381..543W} for full details.

 The strength of the different principle components allows for a clean selection of PSB candidates.  
 In the right panel of Figure \ref{principle_comps} we show as an example PC1 against PC2 for all simulated galaxies between look-back times of 3 and 4 Gyr ($z\approx0.3$), with our simulated PSBs highlighted in red. The PSB spectra are calculated at $t_{\rm PSB}$. While a host of model uncertainties (e.g.\ with stellar population models, dust corrections etc.) are present, the simulated PSBs are clearly distinct from the rest of the population, and the majority would be cleanly selected as a PSB by the PCA method if observed around $t_{\rm PSB}$. 

Other methods of selecting PSBs also exist, and each has a different set of strengths, limitations, and biases \citep{2005MNRAS.357..937G,2008MNRAS.391..700G,2009MNRAS.398..735Y,2016ApJS..224...38A}.
 This test demonstrates, however, that the simulated PSBs, selected as described in Section \ref{howweselect}, have the correct features required to be selected observationally and thus it is reasonable to use these systems to understand the evolutionary processes shaping the observed PSBs.

  \begin{figure} \begin{center}
\includegraphics[width=0.475\textwidth,angle=0,clip,trim=0cm 0cm 0cm 0.0cm]{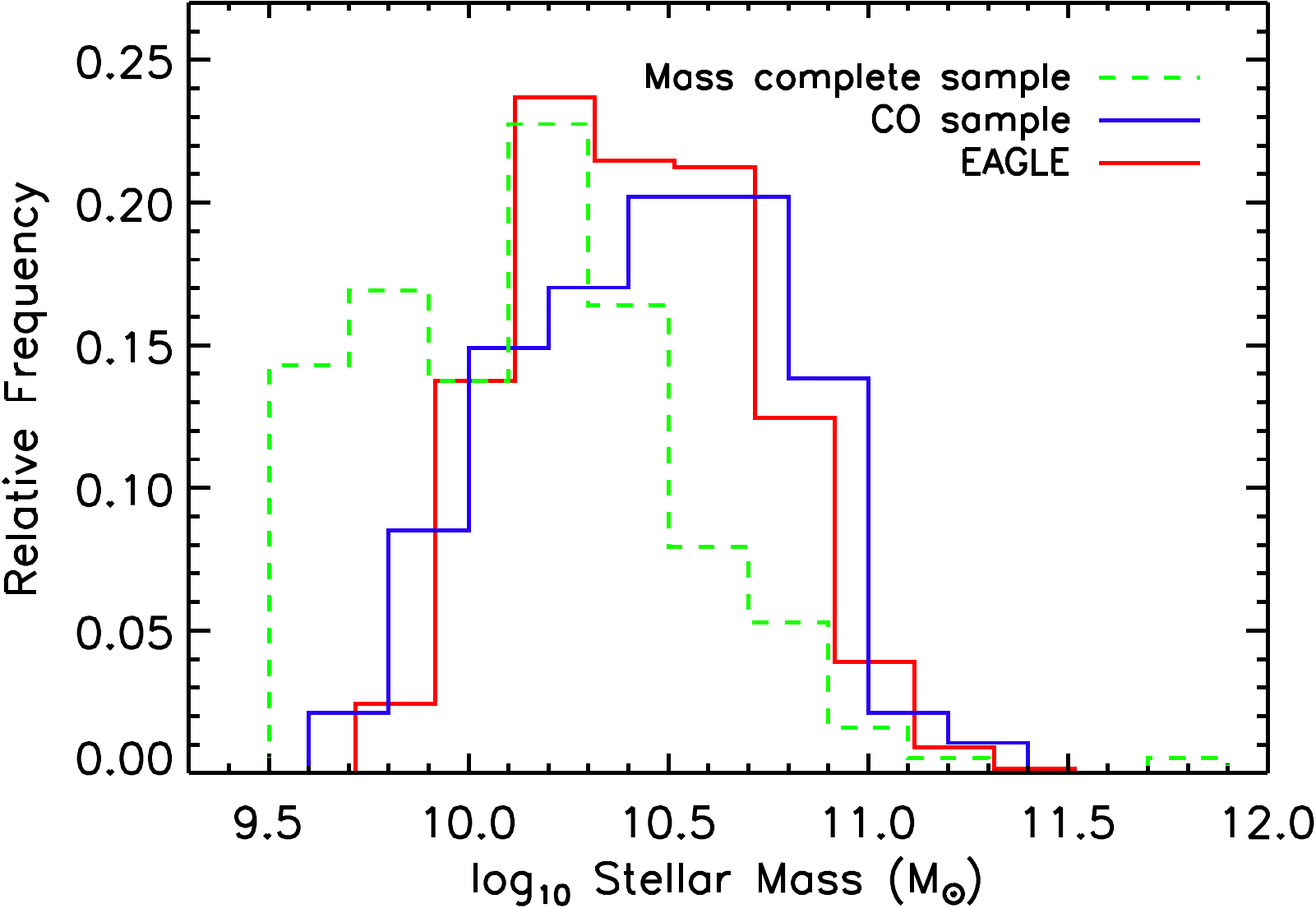}
\caption{Stellar mass distribution of the low redshift simulated PSBs (red, assuming a \protect \citealt{Chabrier2003} IMF) our observational comparison set (blue), and a complete sample of PSBs from \protect \cite{2018MNRAS.477.1708P} (green dashed histogram) . PSBs selected in EAGLE reproduce the observed mass distribution of the gas rich sample well, with a typical PSB mass of $\approx$10$^{10.5}$ \msun.}
\label{mstar_comp}
 \end{center}
 \end{figure}

 \subsubsection{Stellar masses}

Figure \ref{mstar_comp} shows a comparison of the stellar masses of the simulated PSBs (at $t_{\rm PSB}$; shown in red) to the stellar masses of the low redshift molecular gas rich PSB galaxies from our comparison sample (shown in blue) and to the stellar masses of the complete sample of low redshift PSB galaxies from \cite{2018MNRAS.477.1708P} (shown as a dashed green histogram) . The masses calculated from the simulation take into account stellar evolutionary processes such as mass return to the ISM, and include stellar remnants.

The simulated PSB galaxies are higher mass, on average, than those found observationally in complete surveys. This suggests EAGLE may struggle to reproduce these lowest mass PSB galaxies (perhaps due to its limited mass resolution, and the known deficit of cold gas in EAGLEs low mass systems; \citealt{Crain2017}).
Despite this, the simulated PSBs do reproduce the stellar mass distribution of the objects observed in molecular gas well. A Kolmogorov-Smirnov test conducted on the mass distribution of the simulated PSBs and the CO observed galaxies  returns as probability of 0.27, thus we are unable to reject the hypothesis that the stellar masses of the observed and simulated galaxies are drawn from the same distribution.
The EAGLE galaxies thus seem to have a mass distribution similar to our gas rich comparison objects, and thus can act as a fair comparison sample.

Breaking down the stellar mass distribution as a function of time, we find that the stellar masses of the selected PSBs in EAGLE show essentially no redshift dependence. At low redshifts the median stellar mass of the simulated PSBs is $10^{10.3}$ \msun. It increases marginally to $10^{10.5}$ \msun\ at $z=0.7$, and then declines slightly to $10^{10.4}$ \msun\ at $z=1.5$. This is somewhat at odds with the results of several observational studies \citep[e.g.][]{2016MNRAS.463..832W,2018MNRAS.tmp.1823R}, which do find significant evolution in the PSB mass function. This discrepancy is likely due to the anomalously small number of PSBs at low masses within EAGLE.

  \begin{figure} \begin{center}
\includegraphics[width=0.475\textwidth,angle=0,clip,trim=0cm 0cm 0cm 0cm]{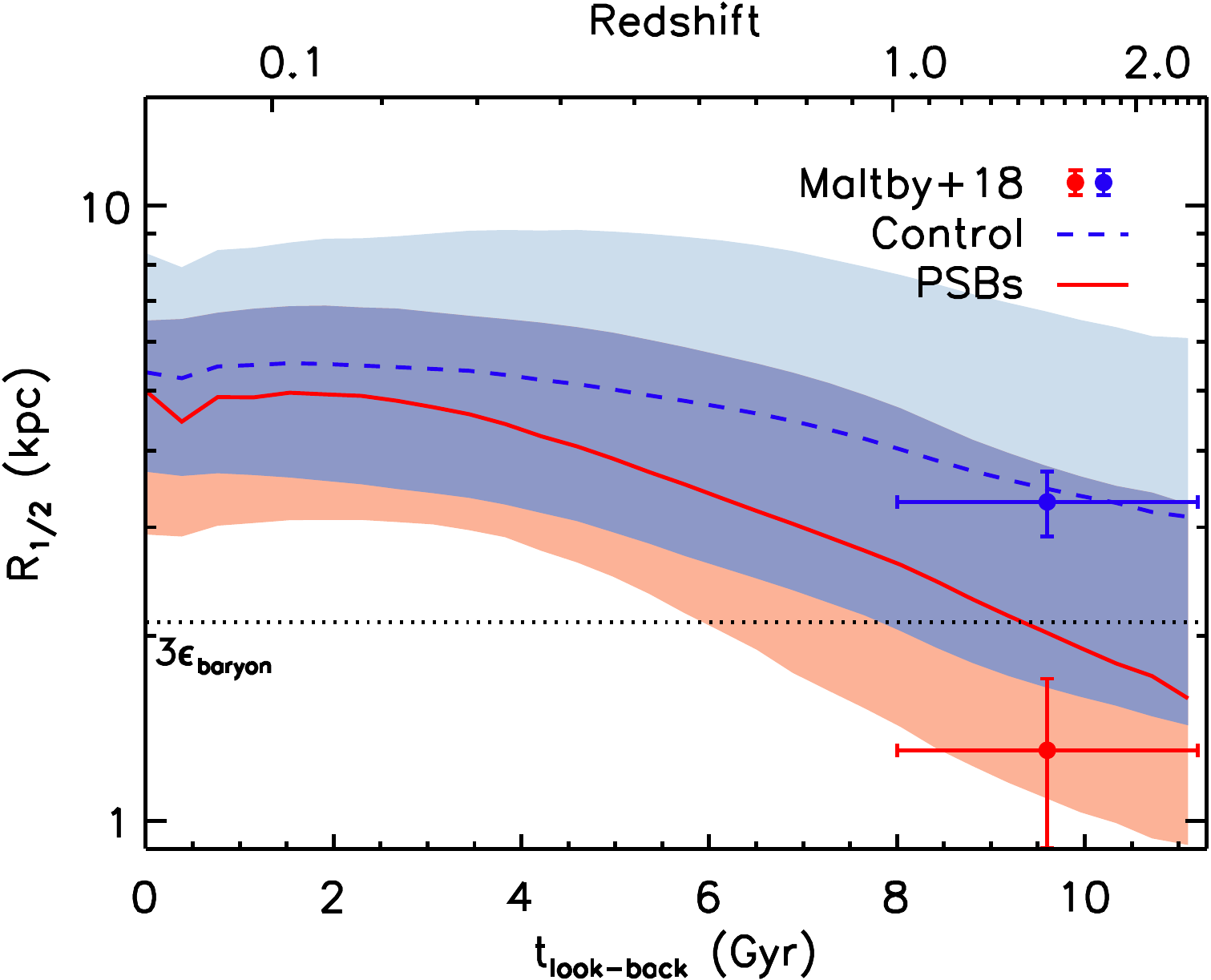}
\caption{Median half-mass size of the stellar component of the simulated PSBs (red curve) and the control sample (blue curve) as a function of look-back time. The shaded regions around the median show the 16th and 84th percentiles. The dotted line shows three times the maximum baryonic smoothing length ($\epsilon_{\rm baryon}$) in the EAGLE simulation, below which galaxy sizes are not well converged. PSBs are smaller than the control galaxies at all redshifts.}
\label{psb_radii}
 \end{center}
 \end{figure}

\subsubsection{PSB sizes}

Figure \ref{psb_radii} shows the median half-mass radius of the stellar component of the simulated PSBs (red line) as a function of time. The shaded region around the median denotes the 16th and 84th percentiles (i.e. the 1$\sigma$ scatter). We also plot the size evolution of the control population (blue dashed curve). One can immediately see that the median sizes of the simulated PSBs are always smaller than those of the control galaxies. This size discrepancy increases with redshift, from a factor of 0.1 dex at $z=0$, up to 0.3 dex at $z=2$. We do, however, note that size of objects in the simulations become uncertain as they approach $\sim$3 times the gravitational softening length ($\epsilon_{\rm baryon}$, shown as a dotted black line in Figure \ref{psb_radii}). As the sizes of the simulated PSBs at high redshift decrease below this limit they should be treated as upper limits.

The strong radius evolution of the PSB population agrees with that found in several recent studies (e.g.\ \citealt{2017MNRAS.472.1401A,2018arXiv180701325M,2018arXiv180901211W}).  \cite{2018arXiv180701325M} in particular studied the size evolution of observed high redshift PSBs and their parent population in two redshift slices. Although their lower redshift sample is not comparable to our simulated galaxies due to a different stellar mass distribution, the higher redshift galaxies can be directly compared, and are shown as red and blue datapoints (with associated errors) in Figure \ref{psb_radii}. These observational size estimates appear consistent with those from the simulation, especially given the different measurement techniques (half mass radii are defined in 3 dimensions in this work, while half light radii were measured as projected on the sky in \citealt{2018arXiv180701325M}) and the resolution limit of our simulations.

\subsubsection{PSB number densities}

\begin{figure} \begin{center}
\includegraphics[width=0.475\textwidth,angle=0,clip,trim=0cm 0cm 0cm 0.0cm]{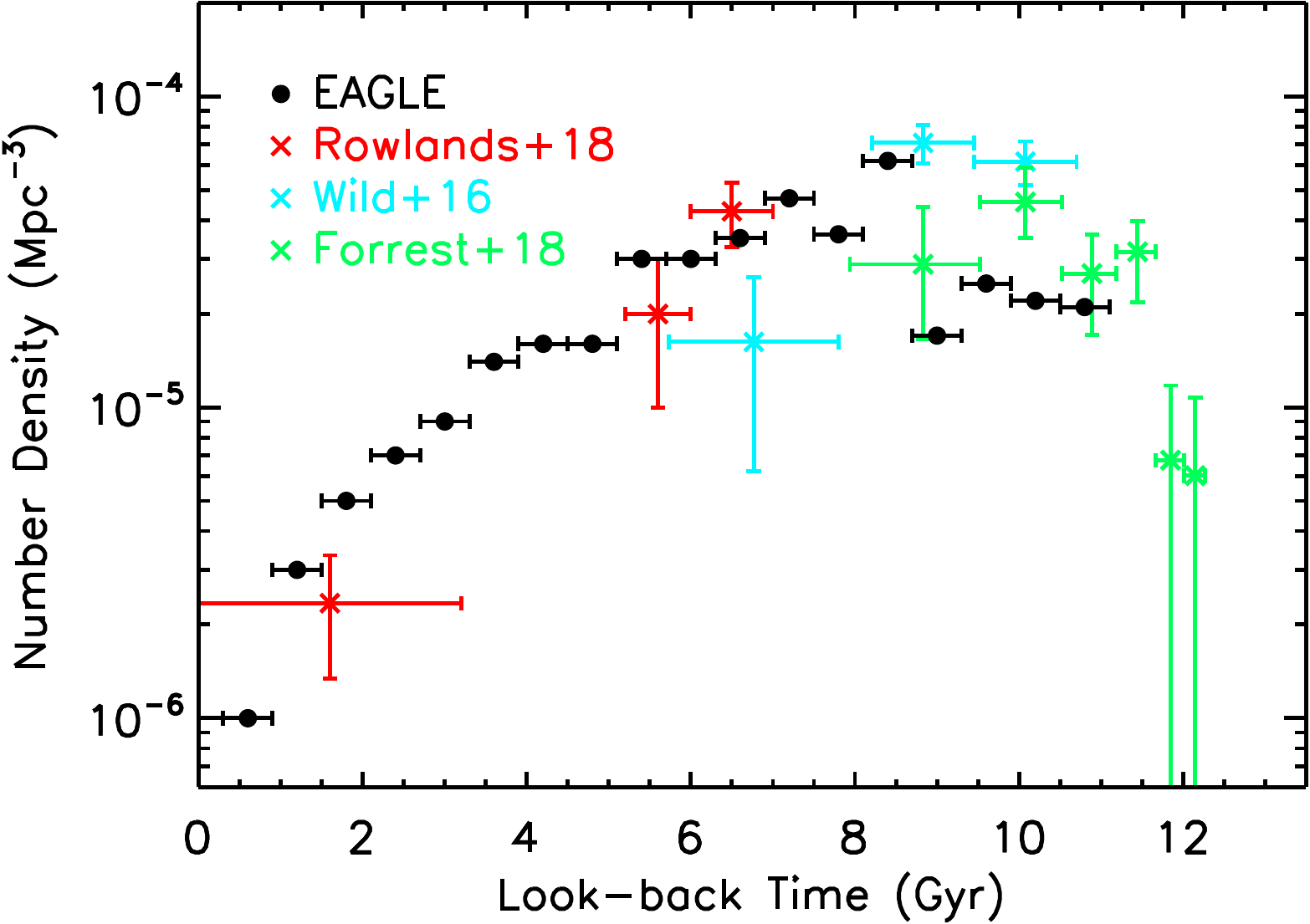}
\caption{Number density of simulated PSBs as a function of look-back time (black points). Error bars indicate the averaging time range (600~Myr). Shown as blue, red, and green crosses are the observational constraints of \protect \cite{2016MNRAS.463..832W,2018MNRAS.473.1168R} and \protect \cite{2018arXiv180703785F} respectively, with their associated observational uncertainties. The simulation well reproduces the observed number densities of PSBs.}
\label{numdens_plott}
 \end{center}
 \end{figure}

Given that we have been able to track the PSB population in the simulation across cosmic time, it is interesting to consider whether the number density of sources we find matches that found in observations. Figure \ref{numdens_plott} shows the number density of simulated PSBs with M$_{\rm *}>10^{10.6}$ \msun\  (where this mass cut was adopted in order to match the selection cuts within the observations, and ensure we are not affected by the lack of low-mass PSBs in EAGLE) plotted as a function of look-back time in 600~Myr bins. We find an increasing trend in the PSB number density out to a look-back time of $\approx$8 Gyr (z$\approx$1), after which it begins to decline. Also shown as coloured data-points with error bars are the measurements of PSB number densities from the studies of \cite{2016MNRAS.463..832W,2018MNRAS.473.1168R} and \cite{2018arXiv180703785F}, which (given their differing methods and selection criteria) agree reasonably well with those extracted from the simulations. This suggests that the observed decline in the number density of high-redshift PSBs may be physical, and not a consequence of high-redshift PSBs being heavily obscured. 

In conclusion, the simulated PSBs selected as described in Section \ref{howweselect} have similar stellar masses, stellar sizes and number densities to observed PSBs, giving us confidence that we are selecting comparable objects. Below we compare the gas properties of these objects, and make predictions for future observations.

  \begin{figure} \begin{center}
\includegraphics[width=0.4\textwidth,angle=0,clip,trim=0cm 0cm 0cm 0.0cm]{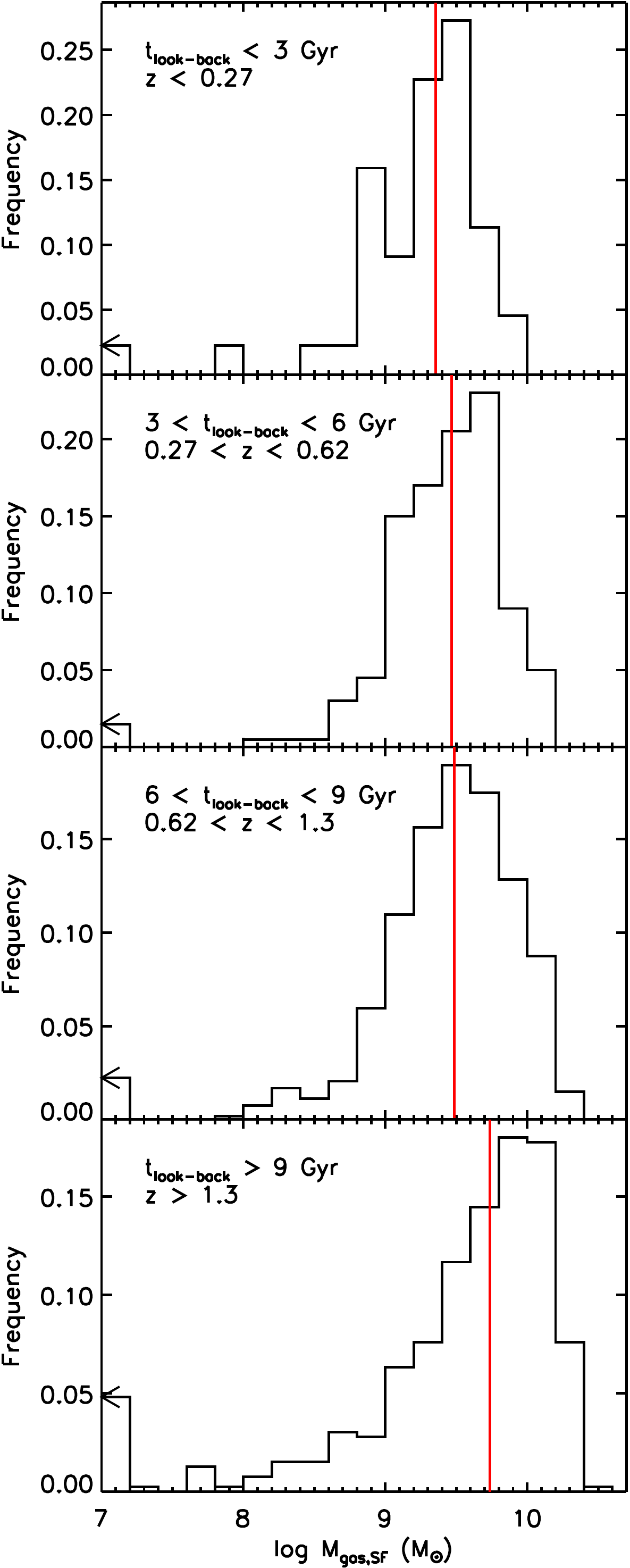}
\caption{Star-forming gas masses of the simulated PSBs evaluated at six times during their 600 Myr PSB phase, split into four redshift bins (as indicated in the legend). The red vertical lines denote the median. Gas-free galaxies have been included in the (otherwise empty) bin at $10^7$\msun, and are indicated by the arrow head. There appears to be little redshift evolution in the star-forming gas mass function of PSB galaxies, at least out to $z\approx1$.}
\label{massfuncs}
 \end{center}
 \end{figure}

\subsection{Cold gas content}
\label{gas_funcs}

In the preceding sections we have shown that we are able to select PSB galaxies from the EAGLE simulations that resemble observed PSB galaxies. We here go on to study the star-forming gas properties of the simulated PSBs, how these match those of observed PSBs, and what these diagnostics tell us about the evolution of these objects. 

Figure \ref{massfuncs} shows the star-forming gas mass of the simulated PSBs, split into four redshift bins. As observed PSBs are captured at different times during their PSB phase, we here sample the gas mass of each simulated object six times after $t_{\rm burst}$, at 100 Myr intervals (the results are not sensitive to this choice of time sampling). Our simulated PSBs have star-forming gas masses between $10^7$ and $10^{10.6}$ \msun. A small fraction ($<$5 per cent) of the PSBs in each redshift bin are totally free of star-forming gas at the mass limit of our simulation, as shown by the histogram bar at 10$^7$ \msun\ with an arrow head. 
 The median gas mass in each redshift bin is shown as a red vertical line. We find very little evolution in the  star-forming gas content of PSB galaxies across cosmic time. The median star-forming gas mass evolves as $\approx(1+z)$, very slowly compared with the observed strong redshift evolution of the total molecular gas content of galaxies over this time period ($\approx(1+z)^{2.5}$; \citealt{2017ApJ...835..120M}).

\begin{figure} \begin{center}
\includegraphics[width=0.475\textwidth,angle=0,clip,trim=0cm 0cm 0cm 0.0cm]{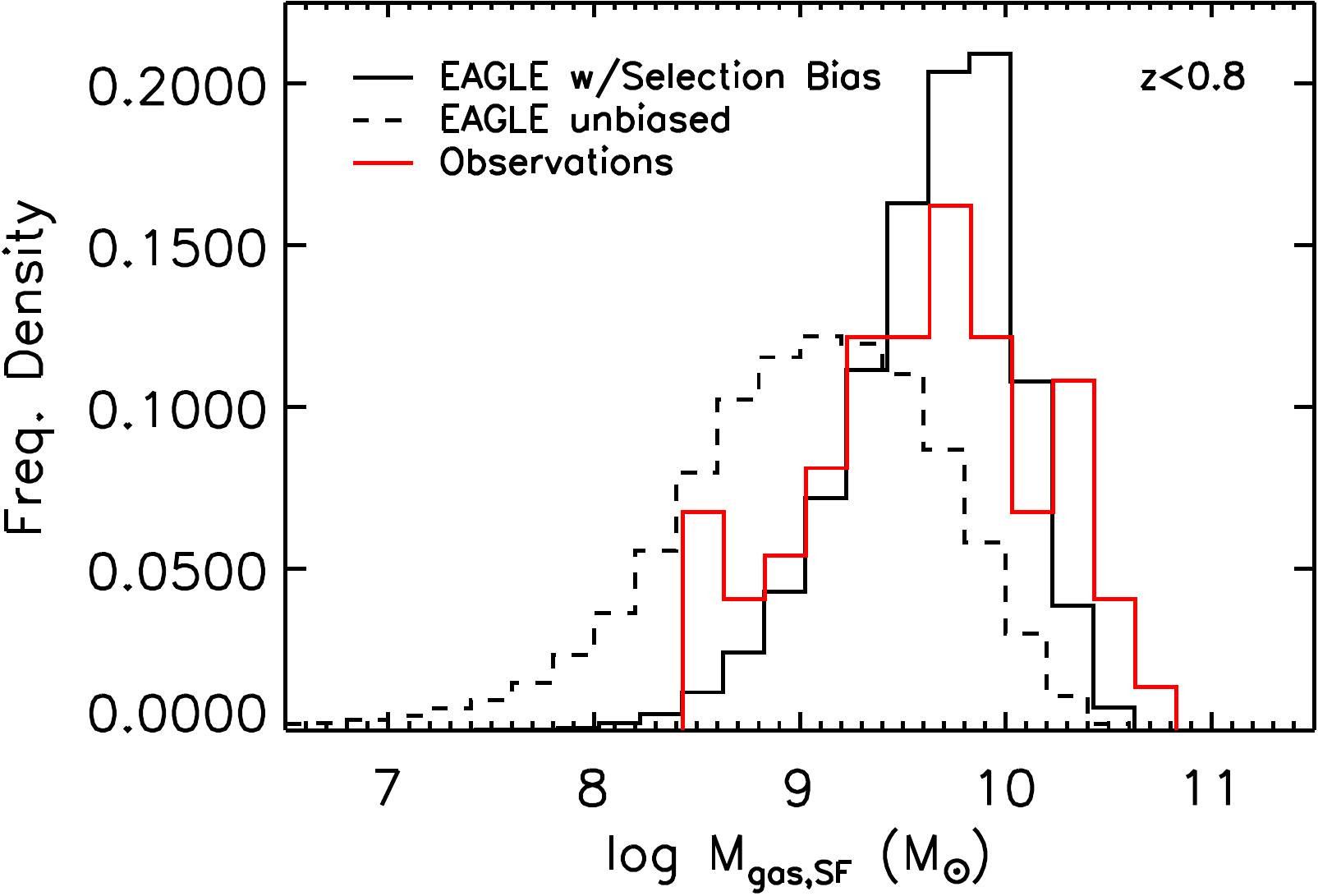}
\caption{Probability density distribution for the star-forming gas mass in our simulated PSBs at $z<0.8$. This is shown directly as found within the simulation (dashed black histogram), and with a simple correction for selection bias as described in the text (solid black histogram). Our simulated galaxies match well the observed star-forming gas mass distribution (red solid histogram).  }
\label{psb_mh2_comp}
 \end{center}
 \end{figure}

Given the above, it is interesting to see how these gas mass estimates compare with the molecular gas observations in our comparison set (described in Section \ref{obs_comp}). 
Figure \ref{psb_mh2_comp} shows a histogram of the frequency density function for the star-forming gas mass in our low redshift simulated PSBs. The gas masses extracted from EAGLE (as described above) are shown as a black dashed histogram, which substantially underestimates the derived gas masses of the combined sample of observed galaxies, shown by the red solid histogram. However, this is likely because the observations suffer from a selection bias, where faint CO emitters can only be detected when they are in nearby objects. The sub-sample of nearby PSBs from \cite{2015MNRAS.448..258R} were all detected, and the molecular gas masses of these objects are consistent with being drawn from the 
simulated distribution. The subsamples of objects from \cite{2015ApJ...801....1F} and \cite{2016ApJ...827..106A} are observed out to much greater distances, however, and as such only the galaxies with more massive cold gas reservoirs are detected.

These observational samples have reasonably similar flux limits, which allows us to make a simple estimate of how incompleteness would affect the derived PDF by inverting the maximum-volume method typically used to correct flux-limited samples \citep{1968ApJ...151..393S}. We do this by assuming the observations reach a fixed CO flux limit of $\approx$3 mK (in main-beam temperature). This allows us to calculate the maximum distance which each of our simulated PSBs would have been detectable at. The correction derived scales with the assumed CO-to-H$_2$ conversion factor ($X_{\rm CO}$), but as these low redshift PSBs are metal-rich we here assume a typical galactic value from \cite{Dickman:1986jz}. Applying this $V_{\rm max}$ correction results in the solid black histogram in Figure \ref{massfuncs}, which (given the uncertainties) agrees well with the observed molecular gas mass PDF.  

We note that limiting the redshift range considered in this analysis to match that probed by the observed sample ($z\approx0.2$) leads to the simulated PSB star-forming gas mass function being badly sampled at the high-mass end (due to the limited volume probed by the simulations as compared to the observations). This lack of the rare, but bright, objects would lead to an $\approx$0.3 dex underestimate of the PSB star-forming gas mass function. Given the negligible star-forming gas mass function evolution found out to $z\approx1$ (see Figure \ref{massfuncs}) we overcome this by extending the redshift range considered out to $z\approx0.8$, providing a better sampling of the PDF, and obtaining good agreement with the observations.

From our selection-bias corrected PDF (in Figure \ref{psb_mh2_comp}) we can estimate the detection rate we would expect the observational surveys to reach if their objects were randomly drawn from our simulated PSB population. 
At the flux completeness limit listed above we would predict a detection rate of $\approx$45 per cent, which (given the simplicity of our approach) matches well with the detection fraction of 53 per cent found in \cite{2015ApJ...801....1F}. This shows that the detection of large gas reservoirs in PSBs, as recently found by a series of authors, is to be expected, and is a natural consequence of the mechanisms causing the PSB episode, which we consider further in Section \ref{quenchmech}.

\subsection{Cold gas evolution}
\label{cold_gas_evol}

Recent work by \cite{2018arXiv180603301F} used a spectral synthesis method to determine the time since the PSB episode began for our observational comparison sample galaxies (those from \citealt{2015ApJ...801....1F,2015MNRAS.448..258R,2016ApJ...827..106A}). They use these times to consider the evolution of the cold gas mass in PSB objects, and determine that a relatively quick evolution is ongoing, with an exponential decay timescale of $\approx$117-230 Myr. They use this as evidence to support a role for active galactic nuceli (AGN) in quenching PSBs. The amount of emission from polycyclic aromatic hydrocarbons in PSBs also seems to be decreasing quickly with post-burst age, supporting a rapid evolution of the ISM during this phase \citep{2018ApJ...855...51S}. Both of theses studies were, however, hampered by lack of knowledge on the initial gas/dust fractions of their objects. 
We thus here check whether the simulated PSBs reproduce the trend seen in \cite{2018arXiv180603301F}, and if a consistent picture emerges for the evolution of the cold gas mass content of galaxies after their PSB episode. 

\begin{figure} \begin{center}
\includegraphics[width=0.475\textwidth,angle=0,clip,trim=0cm 0cm 0cm 0.0cm]{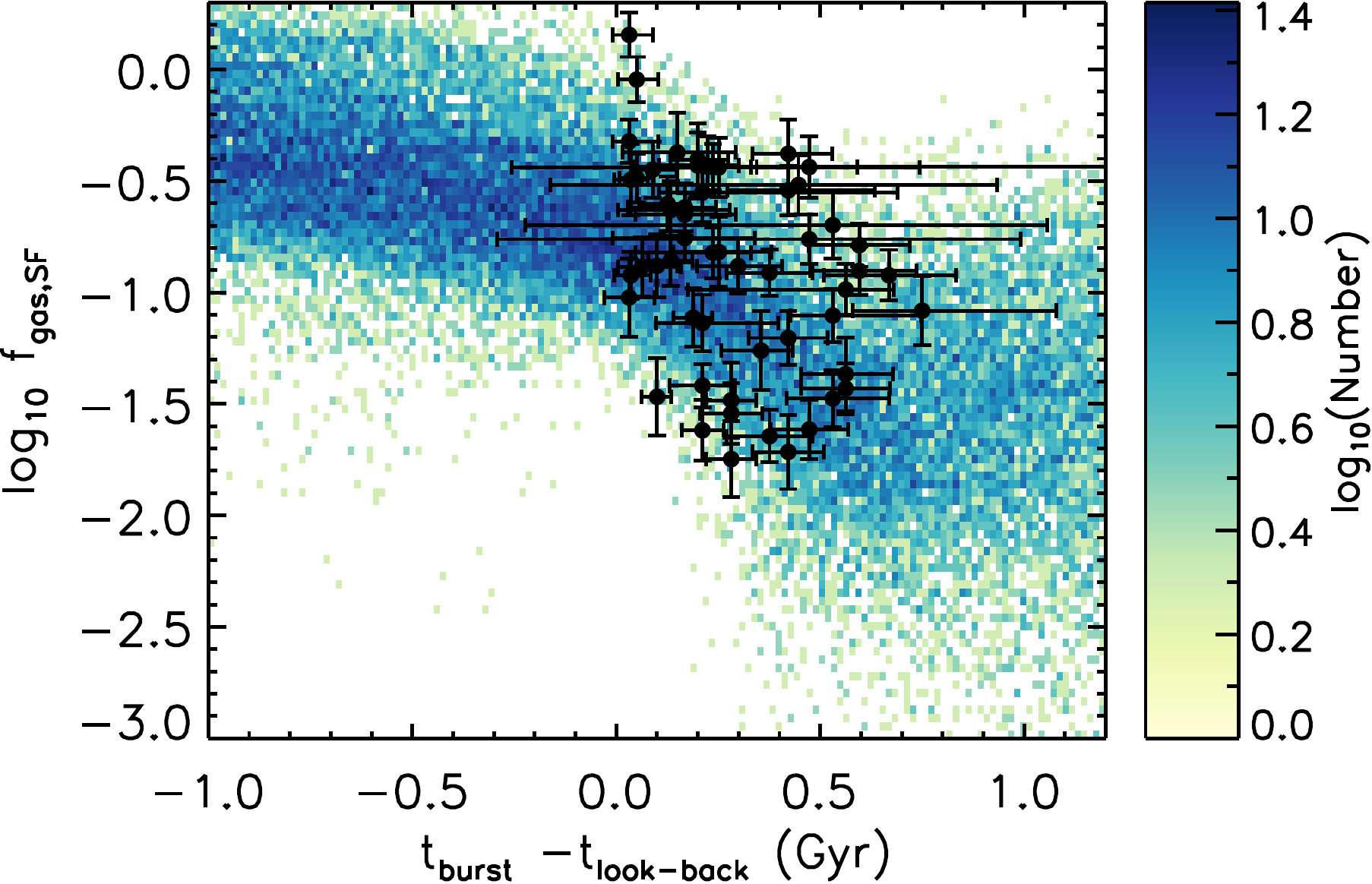}
\caption{Ratio of the star-forming gas mass to stellar mass in the simulated PSBs in the period around their PSB episode (where pixels show the logarithm of the number of points in that bin as colour). Shown as black points with errors are the observational estimates of gas fractions and post starburst event times from \protect \cite{2018arXiv180603301F}. The simulations match the observed decline in the gas-to-stellar mass fraction as a function of time. }
\label{burst_evol}
 \end{center}
 \end{figure}

In Figure \ref{burst_evol} we show (as coloured pixels) the ratio of star-forming gas mass to stellar mass in each of our simulated PSBs at 100 randomly selected times (in the range $-1$ to $+1.2$ Gyr) around $t_{\rm burst}$. 
Also plotted as black points (with their respective error bars) are the {starburst age} measurements for our observational comparison sample \citep{2018arXiv180603301F}. The vast majority of the measurements fall within regions well sampled by our simulated PSBs, suggesting the simulations can well reproduce the observed time evolution of the star-forming gas mass fraction. 

We note that several of the different evolutionary pathways for PSBs highlighted in Pawlik et al.\ submitted are also visible in the late time evolution of PSBs in Figure \ref{burst_evol}, with some galaxies continuing to lose gas mass (consistent with a blue-to-red evolution) while others begin to rebuild their star-forming gas reservoir (a blue-to-blue evolutionary pathway).

 Given the reasonable agreement between the observations and simulations on the post-burst evolution of the gas-to-stellar mass fraction in PSB population, it is interesting to determine if there is an underlying relation, and how individual objects evolve. 
 
\begin{figure} \begin{center}
\includegraphics[width=0.475\textwidth,angle=0,clip,trim=0cm 0cm 0cm 0.0cm]{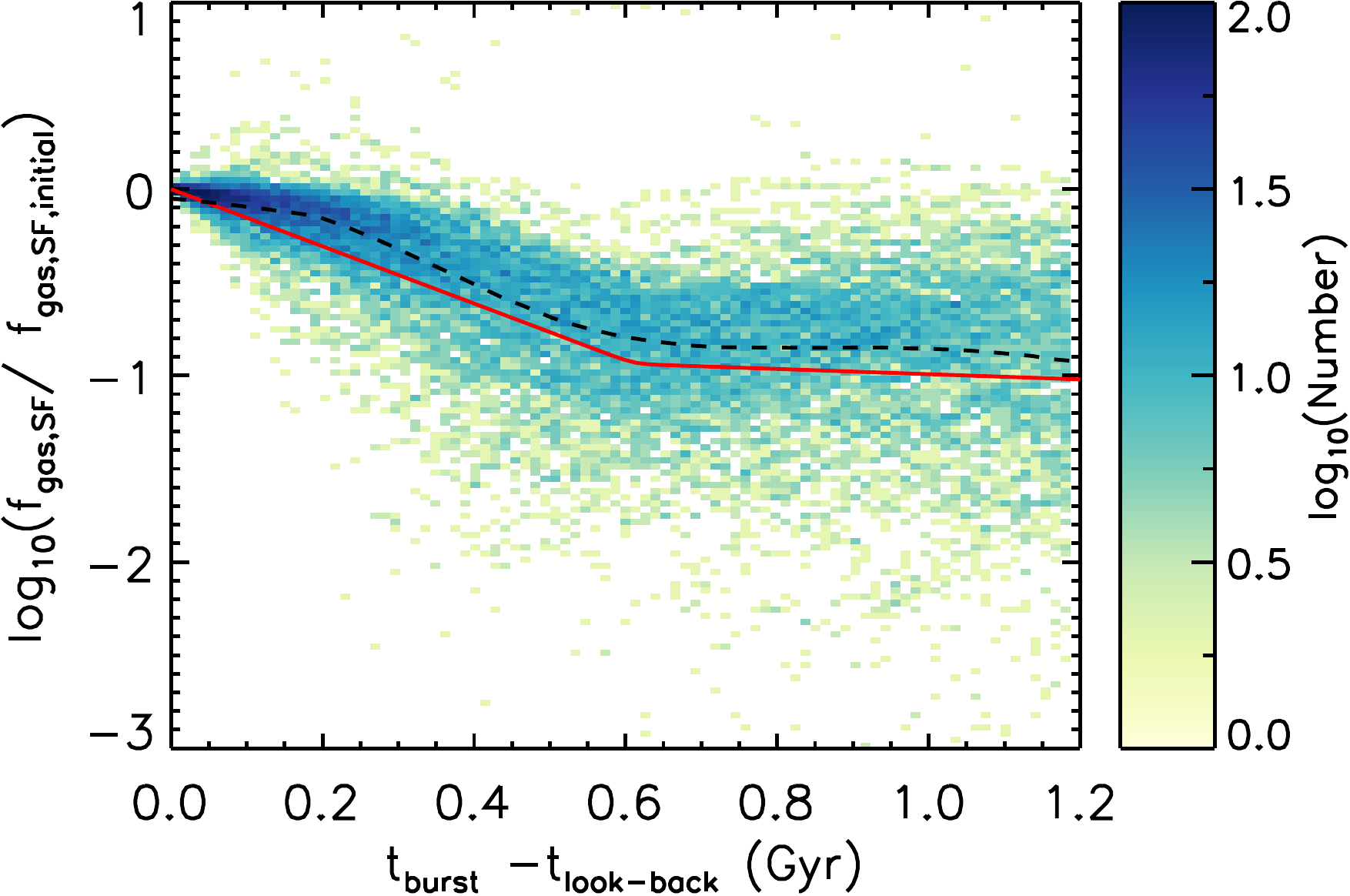}
\caption{As Figure \ref{burst_evol}, but showing the ratio of the star-forming gas mass to stellar mass, normalised by the initial gas fraction in our low redshift simulated PSBs ($z<1$) at $t_{\rm burst}$. Also shown as a black dashed line is the median trend, while the solid red line is the best fit broken exponential with a decay timescale of $\approx$350 Myr and break time of $\approx$600 Myr. }
\label{burst_evol_shift}
 \end{center}
 \end{figure}

 In Figure \ref{burst_evol_shift} we show the evolution of the gas-to-stellar mass fraction as a function of time after the onset of the PSB episode, normalised by the initial gas fraction of each galaxy at $t_{\rm burst}=0$. Normalising the gas-to-stellar mass fraction removes much of the scatter seen in Figure \ref{burst_evol}, revealing a reasonably tight relation between post-burst age and (normalized) gas-to-stellar mass ratio. The median relation is shown as a black dashed line, which declines for $\approx$0.6 Gyr, before flattening. In order to compare the relative timescales for these processes we fit the distribution with a double exponential function (Equation 30 in \citealt{2015ApJ...799..226E}) using the \texttt{IDL} implementation of the \texttt{mpfit} algorithm \citep{2009ASPC..411..251M}. The best-fit is shown as red line in Figure \ref{burst_evol_shift}. 
  We find that the initial decline of the star-forming gas-to-stellar mass fraction is consistent with a fast exponential decay with a characteristic timescale of $\approx300\pm10$ Myr.
  This is similar to the decline time found in both \cite{2010MNRAS.405..933W} and \cite{2015MNRAS.448..258R}, and a little longer than the gas fraction depletion timescale in \cite{2018arXiv180603301F}.
  After  a period of $\approx$620$\pm$50 Myr this decay transitions to a slow decline with a typical timescale of 3$\pm$0.5 Gyr, reasonably consistent with the usual molecular gas depletion timescale of spiral galaxies \citep[$\approx$2 Gyr, e.g.][]{2008AJ....136.2846B}, suggesting whichever violent event is depleting gas early on after the PSB event is typically no longer dominant after this time.  {We note that this change in behaviour is robust to variation in the criteria used to select our simulated PSBs. For instance the same transition at around 600 Myr after $t_{\rm burst}$ is observed if one selects objects with sSFR drops within periods of 400, 600 or 1000 Myr.}

In order to determine what fraction of our galaxies follow the relation shown in Figure \ref{burst_evol_shift} we fitted an exponential function to the first 600 Myr of evolution of each galaxy after their PSB event. A galaxy was considered to be following the median trend if the gradient of their gas fraction evolution with time is the same (within the fitted error) as that for the whole population. Interestingly, around 70 per cent of PSBs follow the median trend, suggesting that Figure \ref{burst_evol_shift} does trace an evolutionary pathway which is followed by many PSBs, despite the many possible mechanisms depleting their gas (see Section \ref{quenchmech}).

\subsection{Star formation efficiency evolution}
\label{sec:sfe_evol}

Many PSBs are thought to have undergone a recent burst of star formation, perhaps thanks to a recent merger or accretion episode. 
Such bursts seem to increase the efficiency of star formation, as observed in some luminous and ultra-luminous infrared galaxies \citep[e.g.][]{2010ApJ...714L.118D,2010MNRAS.407.2091G}. 
Observationally, galaxies in a PSB phase have a longer gas consumption timescale (defined as $T_{\rm dep}\equiv$\,$M_{\rm gas,sf}$/SFR) than is typical in normal spiral galaxies, possibly due to these systems having a low dense gas-fraction \citep{2018arXiv180512132F}. 
In this way PSBs are similar to early-type galaxies, which also host residual low-efficiency star formation \citep[e.g.][]{2011MNRAS.415...61S,2014MNRAS.444.3427D}. Indeed, the longest gas depletion times are found in early-type galaxies which have recently undergone a gas-rich minor merger \citep{2015MNRAS.449.3503D,2018MNRAS.476..122V}.

\begin{figure} \begin{center}
\includegraphics[width=0.475\textwidth,angle=0,clip,trim=0cm 0cm 0cm 0.0cm]{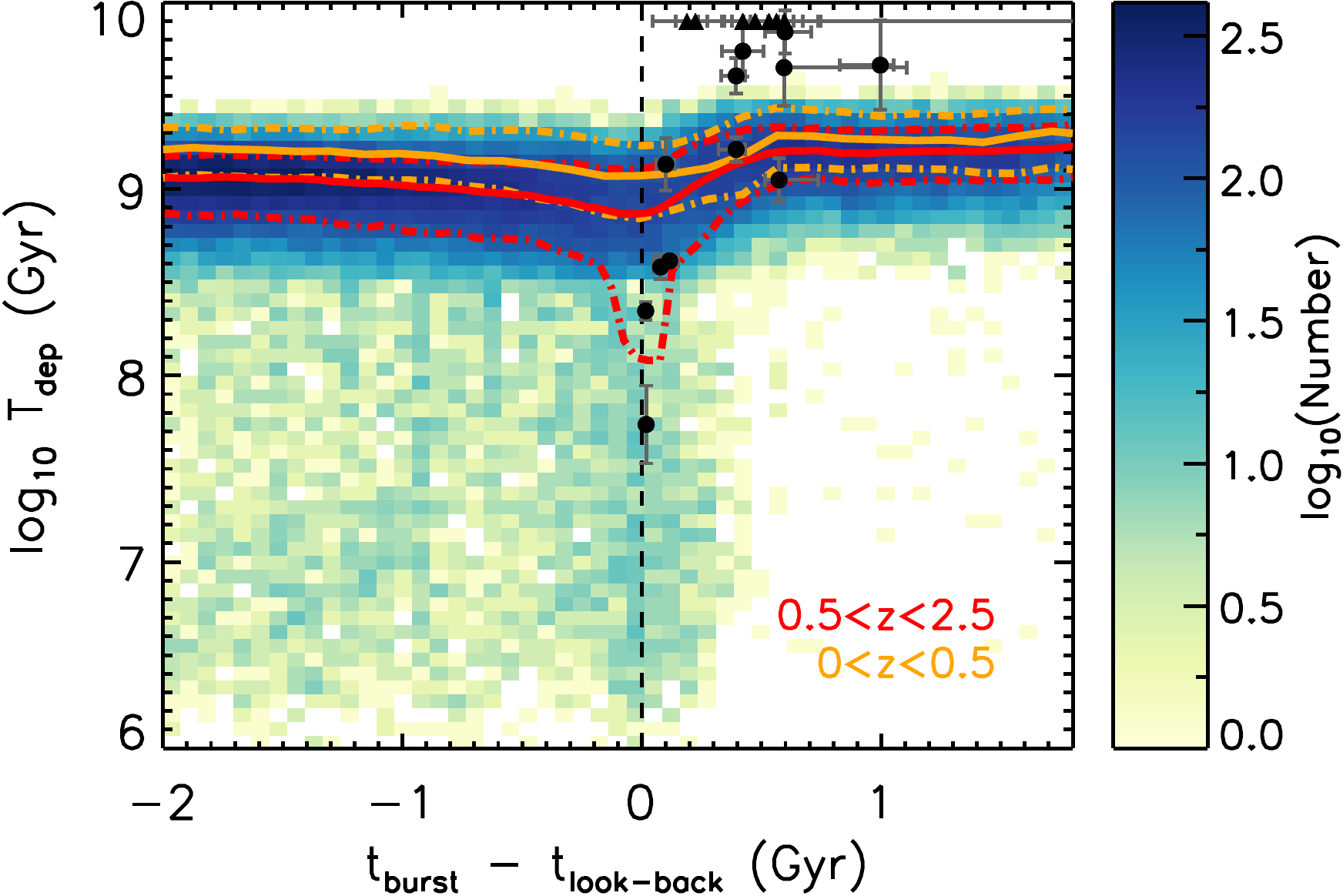}
\caption{Evolution of the gas depletion time during the period around the PSB phase of each of our simulated galaxies. Shown in red and orange curves are the median (solid curve) and 1$\sigma$ scatter (dot-dashed curve) for the high redshift ($z>0.5$) and low redshift ($z<0.5$) PSBs, respectively.  Also shown as black datapoints are the observations of \protect \cite{2015ApJ...801....1F} and \protect \cite{2015MNRAS.448..258R}. The longest depletion times allowed by the EAGLE subgrid star formation model is $\approx10^{9.55}$ Gyr for a high metallicity gas at the lowest ISM densities. A clear evolution to longer depletion times is seen after the PSB event (black dashed line).  }
\label{sfe_evol}
 \end{center}
 \end{figure}

 Due to the finite mass limit and the way star formation is parameterised in the EAGLE simulations there is only a limited range of depletion times which are allowed by the sub-grid model, depending on the gas pressure (or equivalently its density; see Section \ref{data}). 
Despite this, it is interesting to see whether any of the observed trends in depletion time are reproduced in our simulated objects.
In Figure \ref{sfe_evol} we show the depletion time as a function of time relative to $t_{\rm burst}$ for each of our simulated PSBs. The interpolated evolution of the depletion time in each object is sampled every 0.1 Myr, creating a 2D histogram. The median trend, and 16th, and 84th percentile of the distribution are indicated with red/orange solid and dot-dashed curves for PSBs at $z>0.5$ and $z<0.5$, respectively. Where available, we also show the depletion time of the comparison sample objects from \cite{2015ApJ...801....1F} and \cite{2015MNRAS.448..258R}, shown as black points (or black triangles for those with measured depletion times $>10$\,Gyr).

The majority of simulated PSBs form a sequence with fairly constant depletion times (of $\approx$1 - 1.5 Gyr, dependent on redshift) before their PSB incident. A small number of (primarily high redshift) PSBs have efficient starbursts during this time (with depletion times below $\sim$300 Myr), typically caused by ongoing mergers. Around $t_{\rm burst}$ there is a significant increase in the fraction of galaxies undergoing efficient starbursts, with $\approx$16~per~cent of objects in the time period $t_{\rm burst}\pm0.2$\,Gyr having depletion times below $\sim$300 Myr. We note that this fraction is redshift dependent, with only 6 per cent of PSB objects at $z<1$ undergoing such a clear burst of high efficiency star formation. This does not mean that other objects at low redshift did not experience a burst of star formation, but rather that any burst that may be present had a normal SFE. 

After the PSB phase starts, we see a clear shift towards longer depletion times, with the median object doubling its depletion time (from $\approx$1 to $\approx$2 Gyr), and many objects reaching the longest depletion times allowed by the EAGLE subgrid model ($\approx$3.5 Gyr for high metallicity gas at the lowest ISM densities). While this evolution is small compared to that present in the observed PSBs, given the parameterised star formation model of EAGLE this difference is significant, and below we attempt to understand its origin. 

\begin{figure} \begin{center}
\includegraphics[width=0.475\textwidth,angle=0,clip,trim=0cm 0cm 0cm 0.0cm]{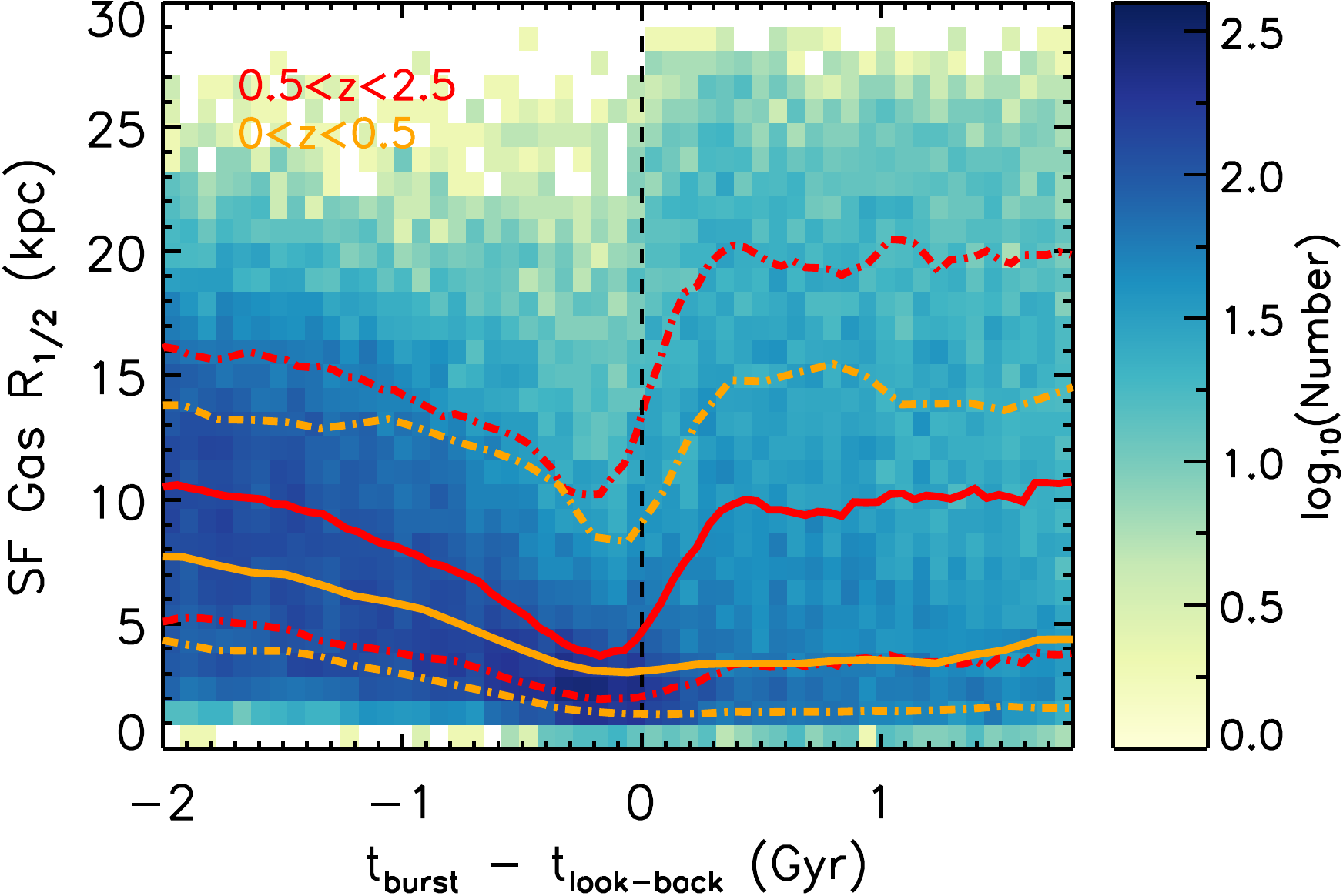}
\caption{As Figure \ref{sfe_evol}, but showing the half-mass size of the star-forming gas in the simulated PSBs, as a function of time around the start of their PSB episode. The median trend is for compaction of the gas reservoir to occur during the 2 Gyr before the PSB episode. After the PSB phase begins the reservoir expands at high redshift, but stays compact at low redshift.}
\label{sfe_evol_size}
 \end{center}
 \end{figure}

Figure \ref{sfe_evol_size} shows the half-mass size of the star-forming gas in the high redshift simulated PSBs, plotted as a function of time around $t_{\rm burst}$. The median trend, and 16th and 84th percentile of the $2.5>z>0.5$ distribution are indicated with red solid and dot-dashed lines respectively, while these quantities for low-redshift ($z<0.5$) galaxies are shown as orange curves. 

The median half-mass size of the gas reservoir in the simulated PSBs decreases from $\approx$10 (proper) kpc to $\approx$3 kpc just before the PSB event. This compaction is likely caused by the inflow of gas \citep[e.g.][]{2015MNRAS.450.2327Z}. This inflow is due to dissipation during a merger in many cases, but can also be due to environmental effects (e.g.\ ram pressure stripping unbinding and heating gas from the outer parts of galaxies). 

After $t_{\rm burst}$ the median size of the gas reservoir remains compact for at least a Gyr, at least at low redshifts. However in $z>0.5$ PSBs the median size of the gas reservoir appears to increase, and after $\approx$0.3 Gyr returns to approximately the same size as before. This is despite an $\approx$5 fold decrease in the gas mass present (see Figure \ref{burst_evol_shift}), implying that the mean gas density is significantly lower. This trend of increasing post-burst half-mass gas size at higher redshift is driven by the higher fraction of increasingly gas rich mergers, which result in disturbed, non-equilibrium gas structures (such as tidal tails). Thus the median size reported in Figure \ref{burst_evol_shift} should not be interpreted as the size of the remaining equilibrium gas structures in these galaxies - which can be significantly more compact.

\begin{figure} \begin{center}
\includegraphics[width=0.475\textwidth,angle=0,clip,trim=0cm 0cm 0cm 0.0cm]{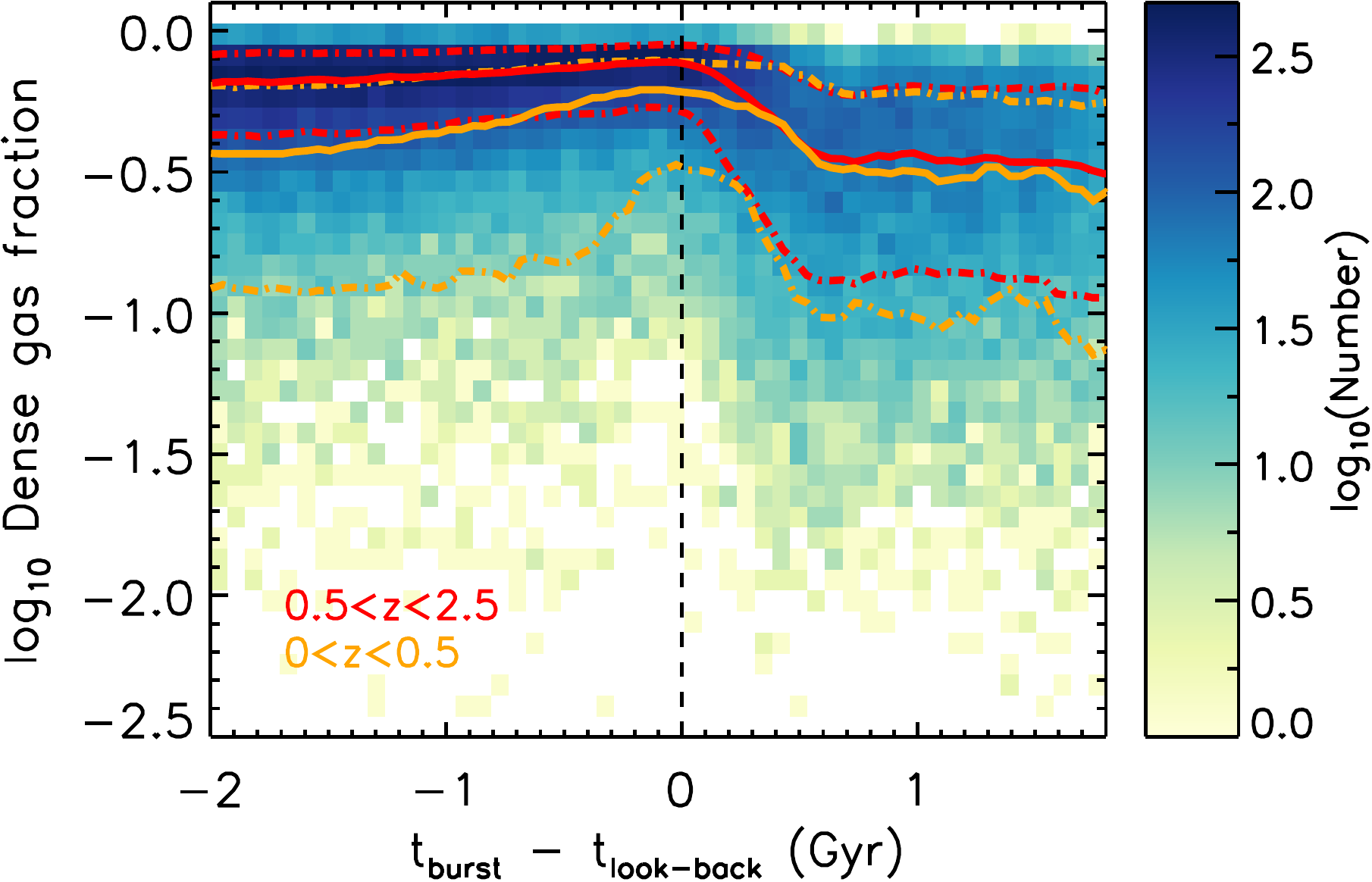}
\caption{As Figure \ref{sfe_evol}, but showing the dense gas fraction (here defined as the fraction of the star-forming gas mass above hydrogen number densities of 1 cm$^{-3}$), as a function of time around the start of the PSB episode of each galaxy. The dense gas fraction increases up until the PSB episode, and then decreases, lowering the SFE.}
\label{sfe_evol_dense}
 \end{center}
 \end{figure}

While the ISM model in EAGLE is simplified (putting star-forming gas on an effective equation of state) it is possible to consider how the relative density of the ISM changes over the PSB episode. Figure \ref{sfe_evol_dense} shows the ``dense gas fraction'' - here defined as the mass fraction of star-forming gas above a hydrogen number density of 1 cm$^{-3}$ (a factor of 10 above the star-formation threshold), as a function of time around $t_{\rm burst}$. The median trend, and 16th and 84th percentiles of the distribution are indicated with red solid and dot-dashed curves respectively for high redshift PSBs, while these quantities for low-redshift ($z<0.5$) galaxies are shown as orange curves.  We caution that these values \textit{cannot} be directly compared to those computed from e.g.\ observed HCN emission \citep{2018arXiv180512132F}. Instead this is simply a summary statistic that can reveal why the SFE in the simulation changes - which may plausibly be the same mechanism (on different scales) as probed by the observations.

The median dense gas fraction is high in the simulated PSBs before their PSB episode, with $>$50 per cent of the gas mass residing in particles with densities $>$1 cm$^{-3}$ when one averages over all redshifts. At low redshift the median dense gas fraction is somewhat lower, with around $30$ of the mass above this density threshold. The dense gas fraction increases slightly at the time each object becomes a PSB, before decreasing markedly, at a similar rate to which the SFE itself declines.  

Figures \ref{sfe_evol_size} and \ref{sfe_evol_dense} paint a consistent picture, where PSB galaxies evolve to have long depletion times due to a lower dense gas fraction (as suggested by \citealt{2018arXiv180512132F}). We predict that future resolved molecular gas observations would find compact molecular reservoirs in low-redshift PSBs, and possibly larger sizes in their higher redshift analogues. Size measurements made using integral-field spectroscopy, or photometry in the rest-frame blue/ultraviolet may also be able to observe the same trend.

\subsection{Quenching mechanisms}
\label{quenchmech}

As discussed in Section \ref{cold_gas_evol}, the initial evolution of the star-forming gas-to-stellar mass ratio in the simulated PSBs is fairly rapid (see Figure \ref{burst_evol_shift}), with an exponential decay timescale of $\approx300$ Myr.
While this is somewhat slower than the $\approx$117-230 Myr found by \cite{2018arXiv180603301F}, it is fast compared to gas depletion times typically observed for unperturbed galaxies. 
Star formation driven winds do not seem to be dominant cause of gas removal in these systems, because the majority of the gas removal seems to be happening hundreds of millions of years after the star formation rate has already dropped. 
Observational evidence suggests that a variety of processes can be important in producing PSB galaxies \citep[e.g.][]{2018arXiv180901211W}.
Within the simulation it is possible to narrow the potential causes of this rapid evolution, and reveal the processes quenching PSBs.

  \begin{figure} \begin{center}
\includegraphics[width=0.475\textwidth,angle=0,clip,trim=0cm 0cm 0cm 0.0cm]{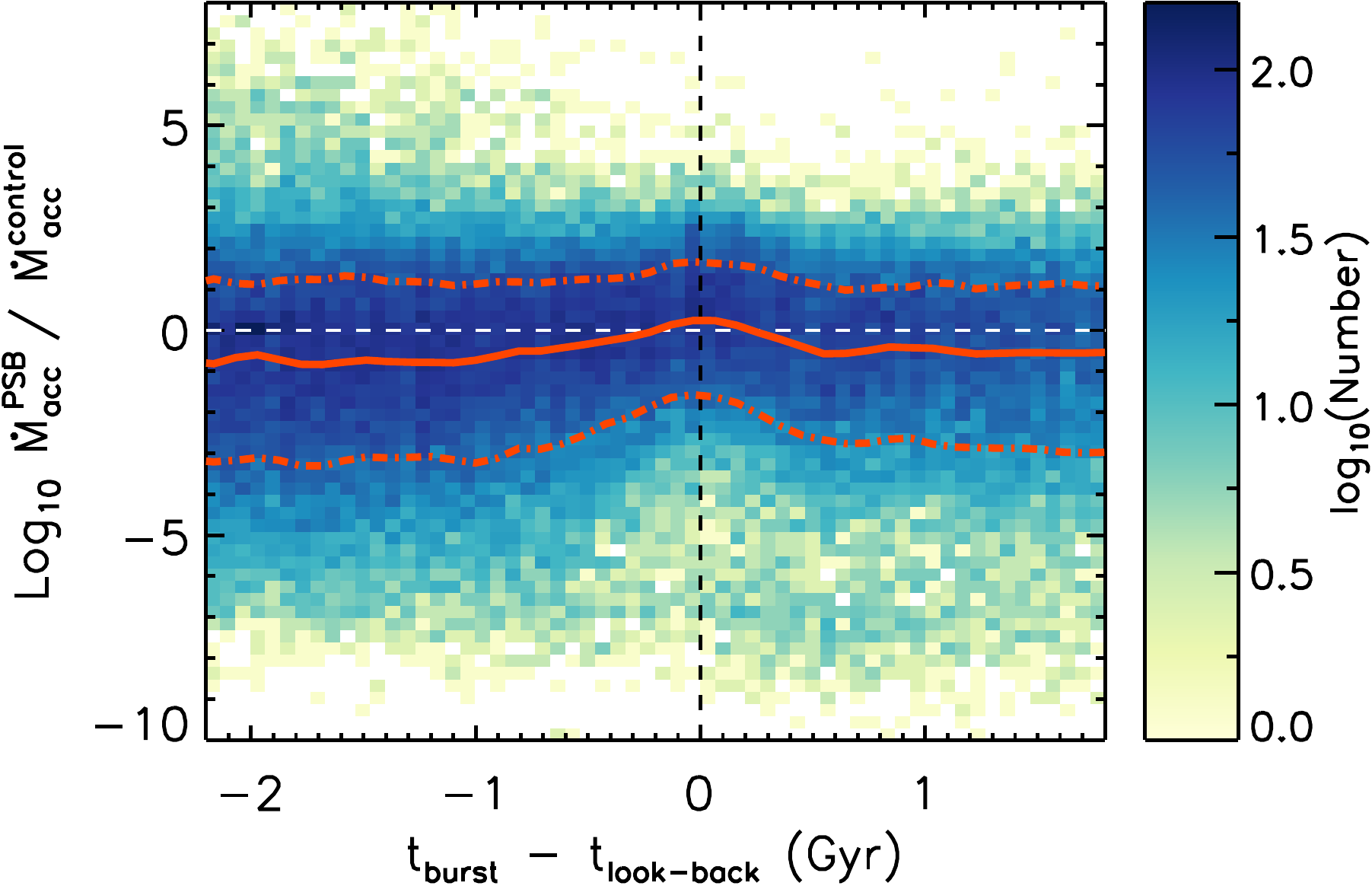}
\caption{As Figure \ref{sfe_evol}, but showing the instantaneous AGN accretion rate of the simulated PSBs, divided by the median accretion rate in every individual objects control galaxies. This is plotted as a function of time around the PSB episode of each galaxy. Low and high redshifts show the same dependance. Only around the beginning of the PSB episode do our PSB canadidates have slightly higher than average accretion rates.}
\label{agn_frac_evol}
 \end{center}
 \end{figure}

 \begin{figure*} \begin{center}
\includegraphics[width=0.7\textwidth,angle=0,clip,trim=0cm 0cm 0cm 0cm]{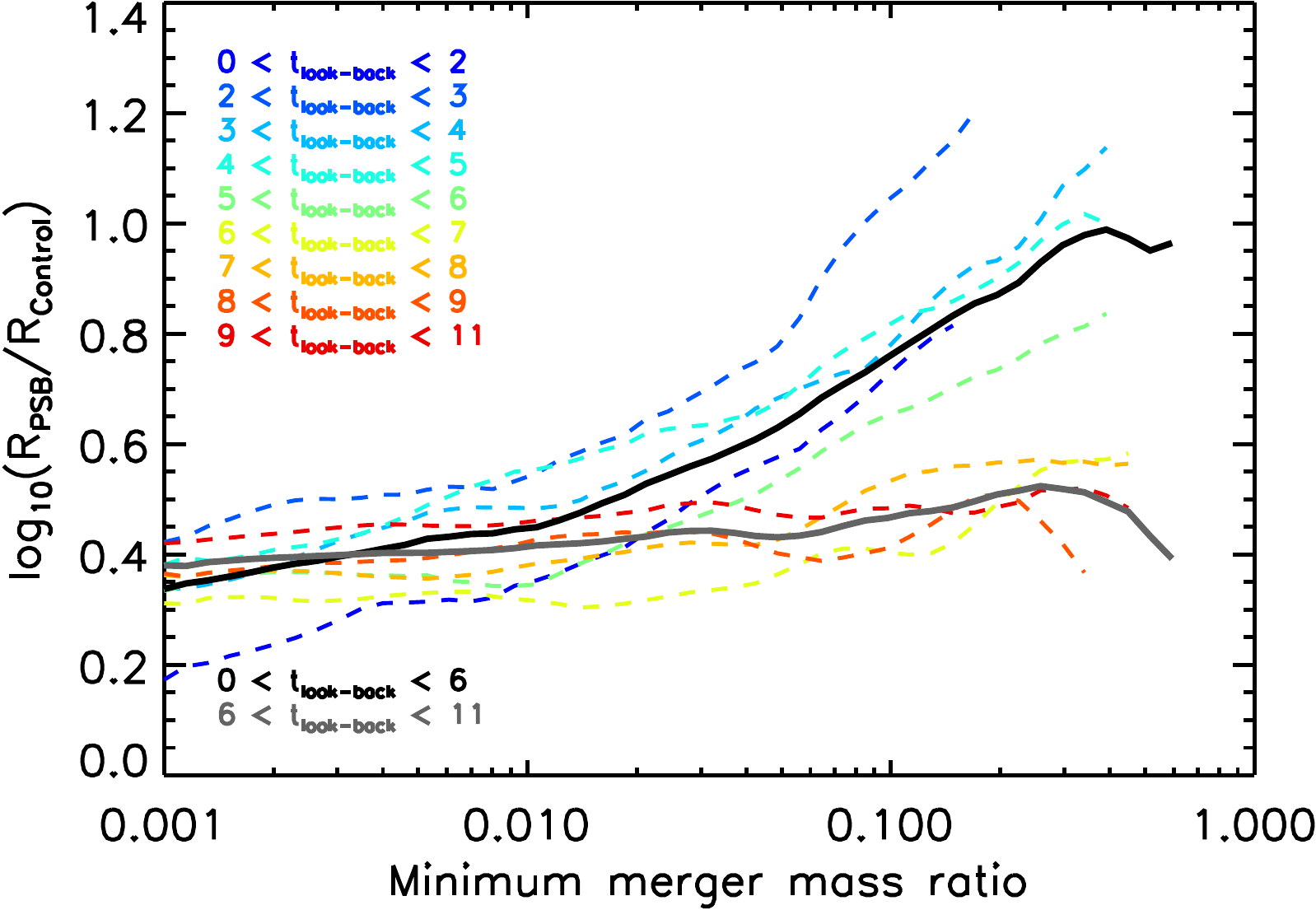}
\caption{Median merger rate in the gigayear surrounding the start of the PSB episode of each of the simulated PSBs, divided by the merger rate of the control galaxies in the same period. This is shown as a function of the minimum merger mass ratio. The relation for the PSB population is shown as coloured curves, split into gigayear intervals. Also shown are summary curves for low (t$_{\rm look-back}<6$ Gyr) and high (t$_{\rm look-back}>6$ Gyr) redshifts. PSB have enhanced merger rates at all times, but major mergers are substantially enhanced (by up to an order of magnitude) at low redshifts. }
\label{merger_rate_comp}
 \end{center}
 \end{figure*}

\subsubsection{AGN activity}

AGN have been implicated in the removal of gas from PSBs, so we here consider their prevalence and importance in quenching our simulated PSBs. 

Non-zero black hole accretion rates are ubiquitous in EAGLE, and hence it is vital to compare the accretion rates of our PSB subsample with the equivalent activity in each individual objects mass-matched control sample.
Although $\approx$1/3 of our PSBs have detectable (here defined as  L$_{\rm bol} > 10^{43}$\,erg/s) bursts of AGN activity during the 600 Myr period of their PSB episode, we find that only a small fraction ($\approx$9 per cent) show any signature of enhanced AGN accretion (here defined as a peak accretion rate $>3\sigma$ above the median level in the control population) during this period. 
A similar fraction of PSBs have enhanced mean accretion rates (when averaging over the whole 600 Myr PSB period) than the control sample objects.
The fraction of objects with enhanced AGN activity increases only slightly (to $\approx$10 per cent) if one also considers the period up to 300 Myr before the PSB phase begins, where starbursts are important. 

This is shown in more detail in Figure \ref{agn_frac_evol}, which shows the black hole accretion rate (computed using Equation 1 in \citealt{2016A&C....15...72M} and evaluated instantaneously at the time of each output snipshot) of the simulated PSBs divided by the median accretion rate in that objects control sample, as a function of time for the period around $t_{\rm burst}$. The median trend, and the 16th, and 84th percentile of the distribution are indicated with red solid and dot-dashed curves respectively. The PSBs span over 7 orders of magnitude in this space, reflecting the strong temporal variability in AGN accretion (see Fig 1 in \citealt{2016A&C....15...72M}). As a population the PSBs have somewhat \textit{lower} accretion rates on average than the control galaxies at almost all times, apart from at $t_{\rm burst}$, when starbursts are occurring. This may be because the black-holes in these systems have yet to enter a rapid growth phase (see \citealt{2018arXiv180508293M}; McAlpine et al.\ in prep).

Overall this suggests the AGN are important in removing the gas from some objects, but cannot be the sole cause, as accretion rates are very low for some objects going through a PSB episode.

 This statistical view of the population is confirmed when one considers the individual histories, such as the one shown in Figure~\ref{ssfr_example}, where there is little to no correlation between the AGN activity and the PSB episode itself. 
{This object does undergo a significant AGN outburst at the end of its post-starburst phase, once the galaxy is already quiescent. This is not a general feature of the population, however, as we find no significant increase of AGN activity in our simulated PSBs for at least 2 Gyrs post-burst when compared with their controls.}
Although EAGLE tracks the accretion rate onto the SMBH sink particles within the simulation, the time resolution with which these are output is much lower than the variability timescale of AGN, and thus it is possible we are missing important energy injection episodes. 
In addition, it is somewhat tricky to determine if any AGN activity present is actually the true cause of gas removal, as it often coincides (and is likely triggered by) the presence of other mechanisms (for instance mergers) which funnel gas towards the SMBH and may also remove it from the system. Given these caveats, the above values should be treated with caution, but still suggest that AGN are not the dominant cause of gas removal in PSB systems in EAGLE.

\subsubsection{Environmental Quenching}
\label{envquench}
The higher incidence of PSB galaxies observed in the densest environments \citep{2018arXiv180511475P} suggests that the processes operating in the
most massive haloes, such as ram pressure stripping \citep{1972ApJ...176....1G} may play a role in these objects. 
Of our simulated galaxies 240/1244 (19 per cent) become a satellite within $\pm$0.5 Gyr of $t_{\rm burst}$. This percentage is consistent with observational estimates of the fraction of PSBs in denser environments \citep[e.g.][]{1996ApJ...466..104Z,2018arXiv180511475P}, and suggests that environment may well be important in triggering PSB episodes in some objects.

In order to quantify the importance of ram-pressure stripping within our simulated PSBs we track the kinematics of the gas particles in our sample objects, and compare these with the motion of the host galaxy during the PSB period. 
We take advantage of the fact that ram-pressure stripping is a directional processes, which preferentially forces gas in the opposite direction to the motion of the galaxy through its host halo.
Within the EAGLE snipshots which cover the PSB episode of each of our objects, we track the particles which were star-forming in one snipshot, and not in the next. 
As long as more than 10 particles meet this criteria we find the direction of the mean vector which describes their displacement, and compare this to the mean velocity vector of the galaxies' centre-of-potential. Objects where the expelled gas preferentially streams away behind the galaxy are good candidates for ram pressure stripping. 

Of our 1244 simulated PSBs 122 (9.8~per~cent) have outflows primarily in the direction opposite to which the galaxy is moving (within $\pm25^{\circ}$).
This fraction is enhanced compared to our control sample galaxies (where $\approx5$\% of objects show these signatures).
Visual inspection of these candidates shows that a small number of false positives are included, typically from major galaxy mergers in their early phases where both galaxies orbit one another and throw out tidal tails of gas behind them. The vast majority, however, have morphologies consistent with ram pressure being the dominant environmental effect. 

As mentioned above, environmental effects, star formation and AGN winds are hard to disentangle fully, and may interact with one another to hasten gas removal \citep[see e.g.][]{2015MNRAS.447..969B}. Despite this, 51 per cent of the newly accreted satellite population that become PSBs show ram-pressure signatures. This suggests that ram-pressure could well be an important mechanism in both triggering PSB episodes, and removing the star-forming gas in PSBs rapidly, at least in dense environments.

\subsubsection{Mergers}
Mergers are common occurrences for field galaxies, and are thought to play a role in triggering the starbursts that initiate the formation of PSBs \citep[e.g.][]{1996ApJ...466..104Z,2004MNRAS.355..713B,2005MNRAS.357..937G,2008ApJ...688..945Y,2009MNRAS.396.1349P,2016MNRAS.456.3032P,2018MNRAS.477.1708P,2018arXiv180701325M}. For our simulated PSBs 358/1244 (29 per cent) have undergone a merger (with stellar mass ratio $>$0.01) within 0.5 Gyr of the onset of their PSB episode. Around 40 per cent of these mergers had a stellar mass ratio $>$0.1, while 10 per cent were 2:1 mergers or higher. The PSB features in these mergers can be caused by the consumption of gas in an extreme starburst, but also by the loss of gas in other ways, for instances in triggered AGN outbursts, or as gas is flung out in tidal tails. Up to $\approx$50\% of the galaxies' gas can be expelled in tails \citep[e.g.][]{2002MNRAS.333..481B}, which either escape, or eventually fall back onto the remnant galaxy. The infall of these tidal tails returns gas on long timescales, and is likely important in the blue-to-blue pathway some PSBs appear to follow (e.g.\ \citealt{2018MNRAS.477.1708P}, Pawlik et al.\ submitted).

In Figure \ref{merger_rate_comp} we plot the merger rate (within $\pm$0.5\,Gyr of $t_{\rm PSB}$) for our simulated PSBs, divided by the merger rate in the control galaxies as a function of the minimum merger ratio. We split the PSB population by look-back time in 1~Gyr intervals (shown as coloured dashed lines), and also show summary curves for low- and high-redshift (solid black and grey lines). 
At all redshifts the merger rate in the PSB population is enhanced relative to the control sample.  At low redshift $z\ltsimeq0.6$ the largest mergers are an order of magnitude more common in the PSB population than the controls. 
At higher redshift, where mergers of all types are more common, the PSB population shows an almost uniform enhancement in the derived merger rate.

Overall we thus conclude that, within EAGLE, major mergers are more important in creating PSBs at low redshift while more numerous minor mergers dominate at high redshift. This is in contrast to the results of \cite{2018arXiv180701325M}, who conclude that major mergers are most important at high redshifts.  The difference between that study and this one is likely partly due to selection, as their lower redshift ($0.5<z<1$) sample is of significantly lower mass than we are able to probe here.

\subsubsection{Summary}

\begin{figure*} \begin{center}
\includegraphics[width=0.9\textwidth,angle=0,clip,trim=0cm 0cm 0cm 0cm]{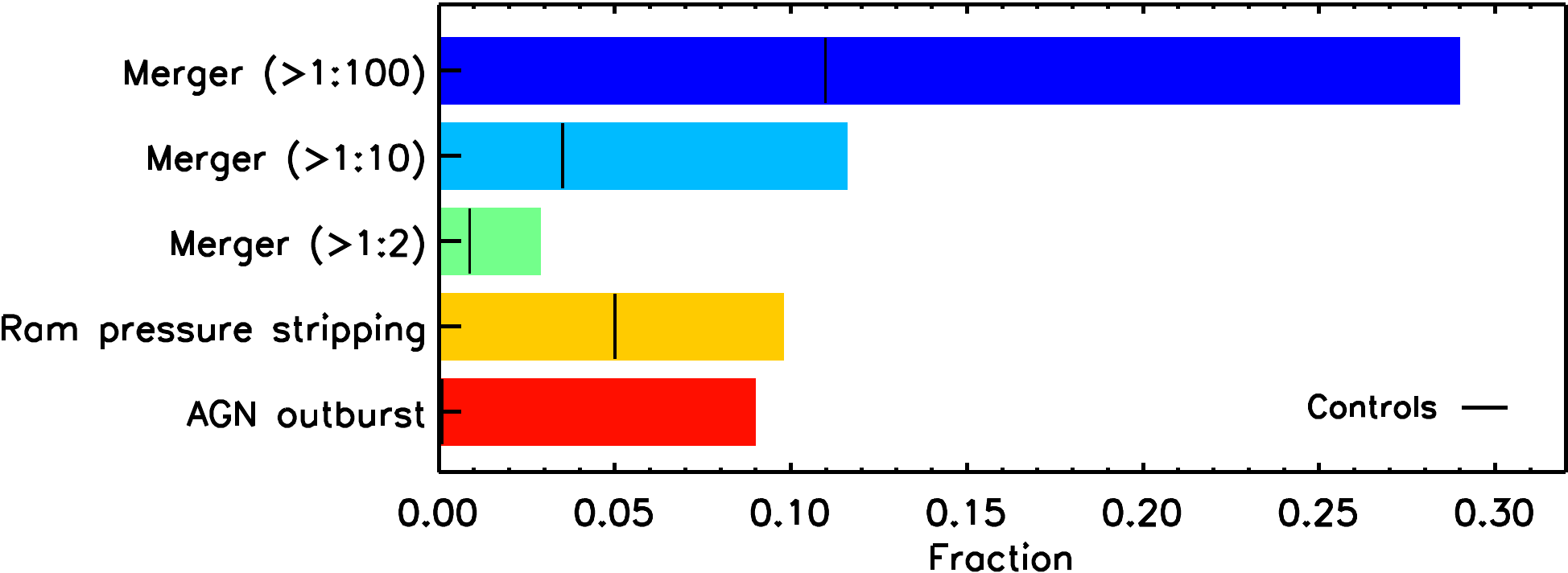}
\caption{The fraction of simulated PSBs which have clearly undergone a given process (described on the $y$-axis) in the $\pm$0.5 Gyr around $t_{\rm PSB}$. The fraction of the control galaxies that underwent these processes in the same period is indicated with a black line on each bar. AGN outbursts are defined a being $>3\sigma$ above the median of the control sample. PSBs appear to be formed by a wide variety of processes, with micro-mergers being the most important.}
\label{barplot}
 \end{center}
 \end{figure*}

In summary, we find that a wide variety of mechanisms are involved in causing galaxies to enter a PSB phase, and for removing gas and suppressing star formation during that phase. The histogram presented in Figure \ref{barplot} summarises the different mechanisms, showing the fraction of all our PSBs that are likely being affected by each process as a histogram bar. The fraction of the control galaxies undergoing this process is indicated with a black line on each bar. The categories are not exclusive, and do not sum to 100 per cent, but give an idea of the variety of processes affecting PSBs, and their relative importance.

\subsection{Gas Kinematics}
\label{kappa_sec}
Observationally, many studies of the cold ISM use signs of disturbance to disentangle objects undergoing secular evolutionary processes from those experiencing more disruptive events. Only a limited number of resolved kinematic measurements have been made on gas-rich PSB galaxies \citep[see e.g.][]{2009MNRAS.396.1349P,2012MNRAS.420..672S,2017A&A...600A..80K}, however, with ALMA and the new generation of large optical integral field unit surveys it will be possible to expand the number of observed objects greatly. We here aim to predict the degree of disturbance one will find in these objects, and include mock observations for comparison to future observations. 

One way of classifying the degree of kinematic disturbance in a simulated gas disc is with $\kappa$, the fraction of kinetic energy invested in ordered rotation \citep[e.g.][]{2012MNRAS.423.1544S}. This is defined as 

\begin{equation}
\kappa=K_{\rm rot}/K_{\rm tot},
\end{equation}
where $K_{\rm rot}$ is the kinetic energy in ordered rotation, defined as
\begin{equation}
K_{\rm rot} = \sum_{i=1}^N\frac{1}{2}m_i\left(\frac{j_{z,i}}{R_i}\right)^2,
\end{equation}
and $K_{\rm tot}$ is the total kinetic energy
\begin{equation}
K_{\rm tot} = \sum_{i=1}^N\frac{1}{2}m_i v_i^2,
\end{equation}
where $N$ is the total number of star-forming gas particles within a given aperture, $m_i$, $v_i$ and $R_i$ are the mass, total velocity and radius (from the centre of potential) of the $i$th particle and $j_{z,i}$ is its angular momentum in the direction of the angular momentum vector. 

\cite{2012MNRAS.423.1544S} used this parameter as a quantitative measure of simulated galaxy \textit{stellar} morphology. Here we simply extend this concept to consider the gaseous components of galaxies (denoting this parameter $\kappa_{\rm gas}$). This parameter is unfortunately hard to estimate observationally as it requires full 6D information (although advanced machine learning techniques may make this possible; see Dawson et al.\ in preparation). Nonetheless, $\kappa_{\rm gas}$ provides a simple way to classify kinematic disturbance, that correlates well with classifications obtained by inspecting the velocity fields by eye. 

\begin{figure} \begin{center}
\includegraphics[width=0.475\textwidth,angle=0,clip,trim=0cm 0cm 0cm 0cm]{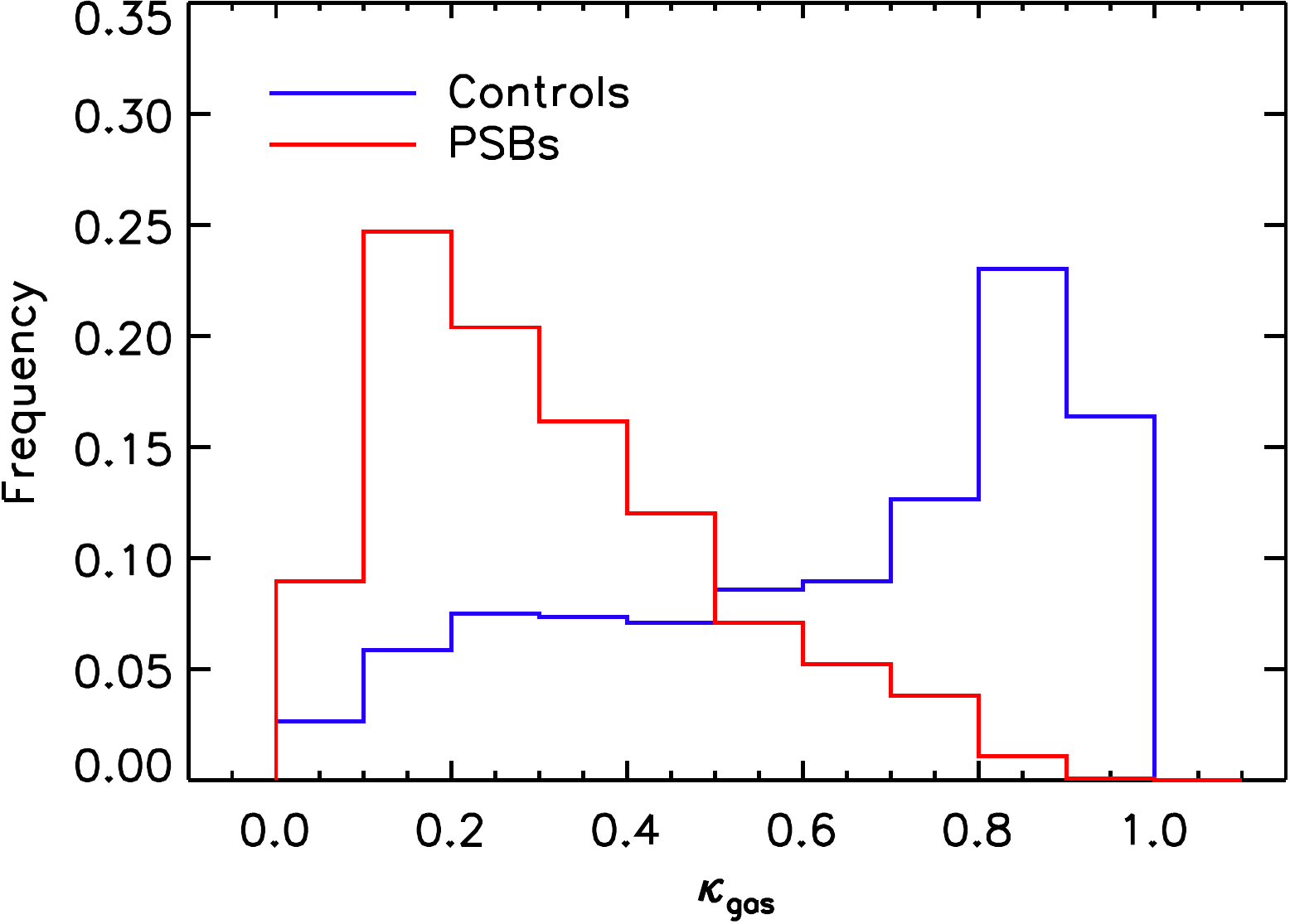}
\caption{The fraction of kinetic energy in the gas phase which has been invested in ordered rotation ($\kappa_{\rm gas}$) for the $z<0.5$ PSBs (red) and their control sample (blue). The gas in the PSBs is substantially more disturbed.}
\label{kappa}
 \end{center}
 \end{figure}

Figure \ref{kappa} shows the distribution of $\kappa_{\rm gas}$ for the simulated PSBs in red, and the control galaxies in blue. 
The two populations show very different distributions. The control sample typically has quite relaxed gas discs, while the gas discs in the PSBs are typically very disturbed, with a low fraction of their total kinetic energy invested in ordered rotation. In this figure we concentrate on low redshift PSBs ($z<0.5$), as these are the most likely to be probed observationally. At higher redshift the control sample objects are more disturbed on average, but this clear dichotomy remains. If one were to break up this histogram as a function of time then the PSB population shows some signatures of relaxation, with the average $\kappa_{\rm gas}$ increasing by $\approx$0.1 over the $\approx$600 Myr PSB phase. The distribution remains broad, however, and dominated by object-to-object scatter, consistent with fairly long relaxation times \citep{2015MNRAS.451.3269V,2016MNRAS.457..272D}.

One subtlety of this parameter is that it requires knowledge of the primary axis of each system. If one uses the angular momentum of the stars to define this, then one obtains a measure of the fraction of kinetic energy invested in ordered rotation \textit{in the same plane as the stars}. In this case a perfectly ordered rotating polar structure, for example, would have a very low $\kappa_{\rm gas}$. Alternatively if one uses the gas itself to define the primary axis, then one gets a measure of how disturbed the gas is, without taking into account the stellar morphology. In Figure \ref{kappa} we show $\kappa_{\rm gas}$ defined with respect to the gaseous kinematic major axis. However, we repeated this analysis using the stellar distribution to define the primary axis then the picture changes little, suggesting the gas in these systems is often very disturbed, and not (yet) settled into regular rotating structures around \textit{any} axis. 

   \begin{figure*} \begin{center}
    \begin{subfigure}[t]{1\textwidth}
            \centering
\includegraphics[height=4.85cm,angle=0,clip,trim=0cm 0cm 0cm 0cm]{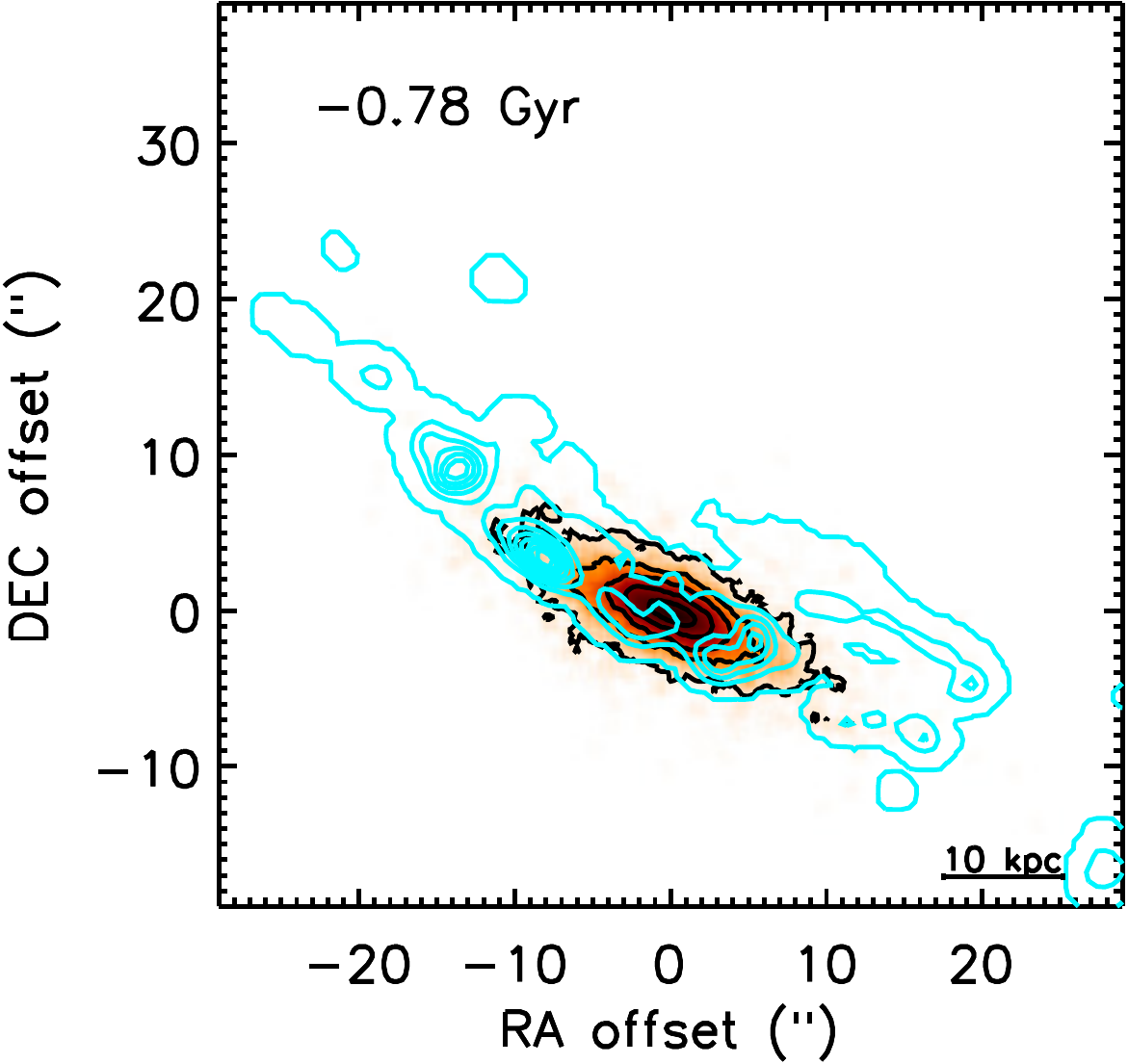}\includegraphics[height=4.85cm,angle=0,clip,trim=2.2cm 0cm 0cm 0cm]{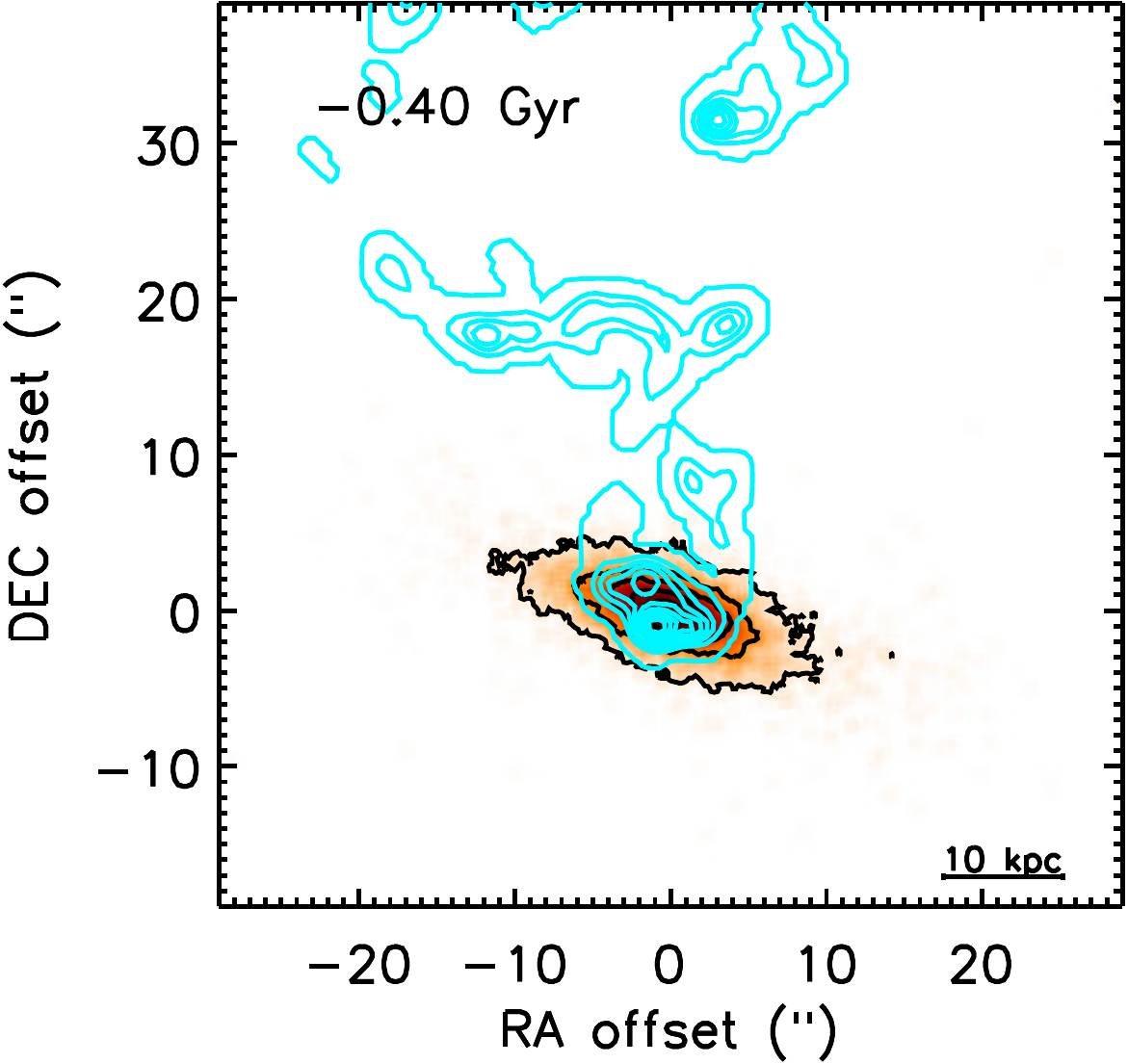}\includegraphics[height=4.85cm,angle=0,clip,trim=2.2cm 0cm 0cm 0cm]{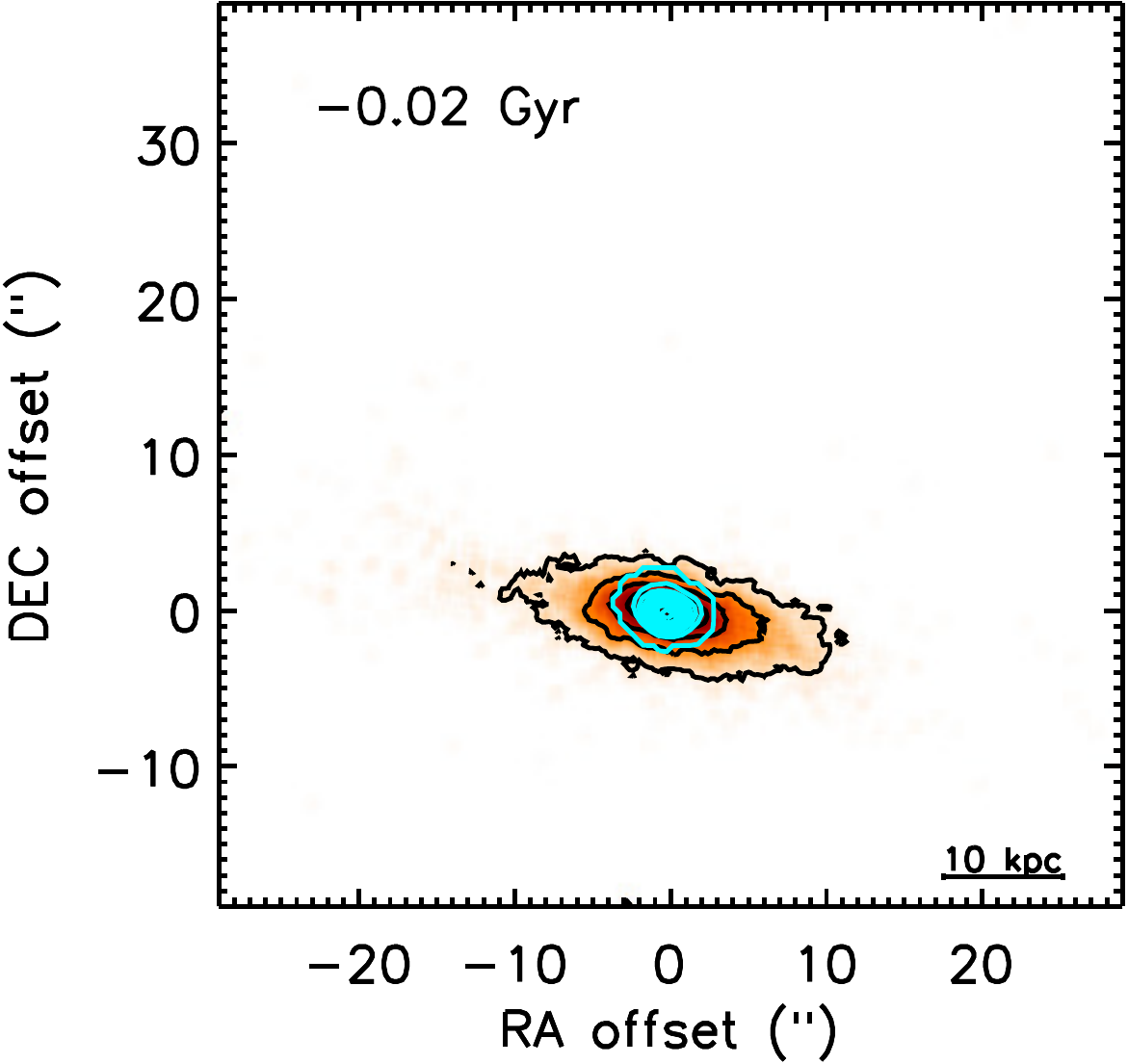}\includegraphics[height=4.85cm,angle=0,clip,trim=2.2cm 0cm 0cm 0cm]{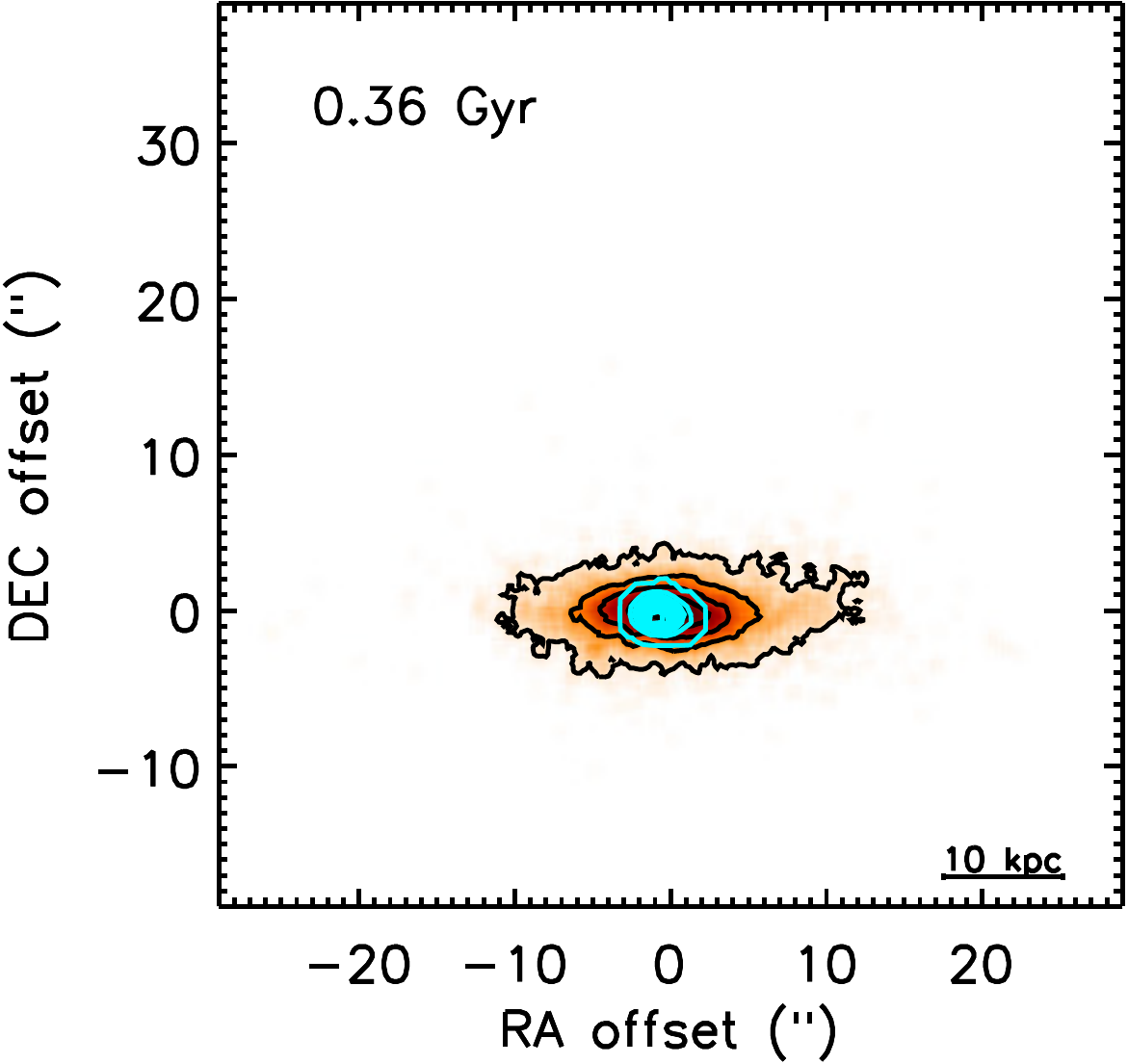}
 \caption{Mock observations of PSB29; post-starburst episode induced by ram-pressure stripping.}
    \end{subfigure}\vspace{0.5cm}
        \begin{subfigure}[t]{1\textwidth}
            \centering
 \includegraphics[height=4.85cm,angle=0,clip,trim=0cm 0cm 0cm 0cm]{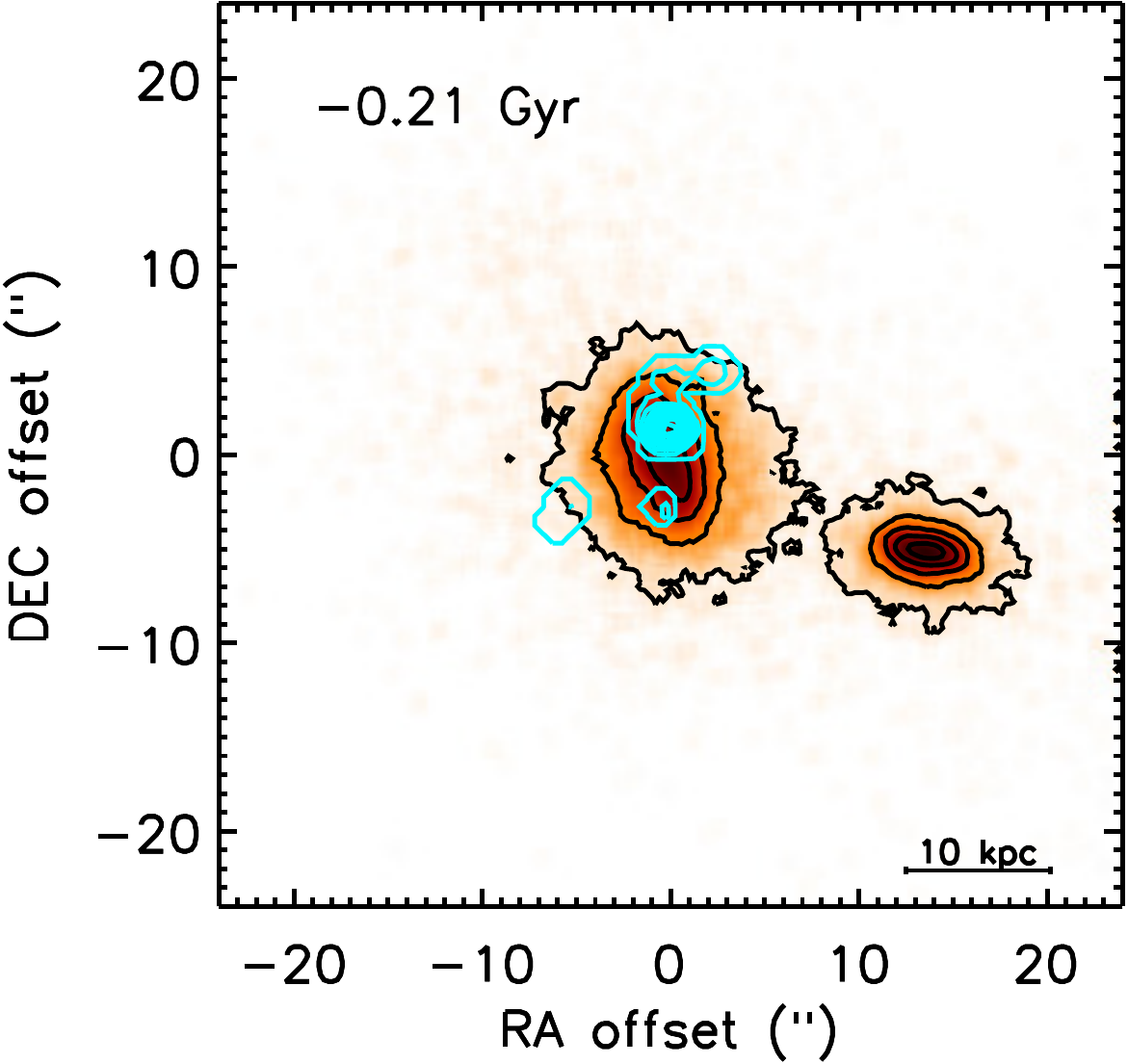}\includegraphics[height=4.85cm,angle=0,clip,trim=2.2cm 0cm 0cm 0cm]{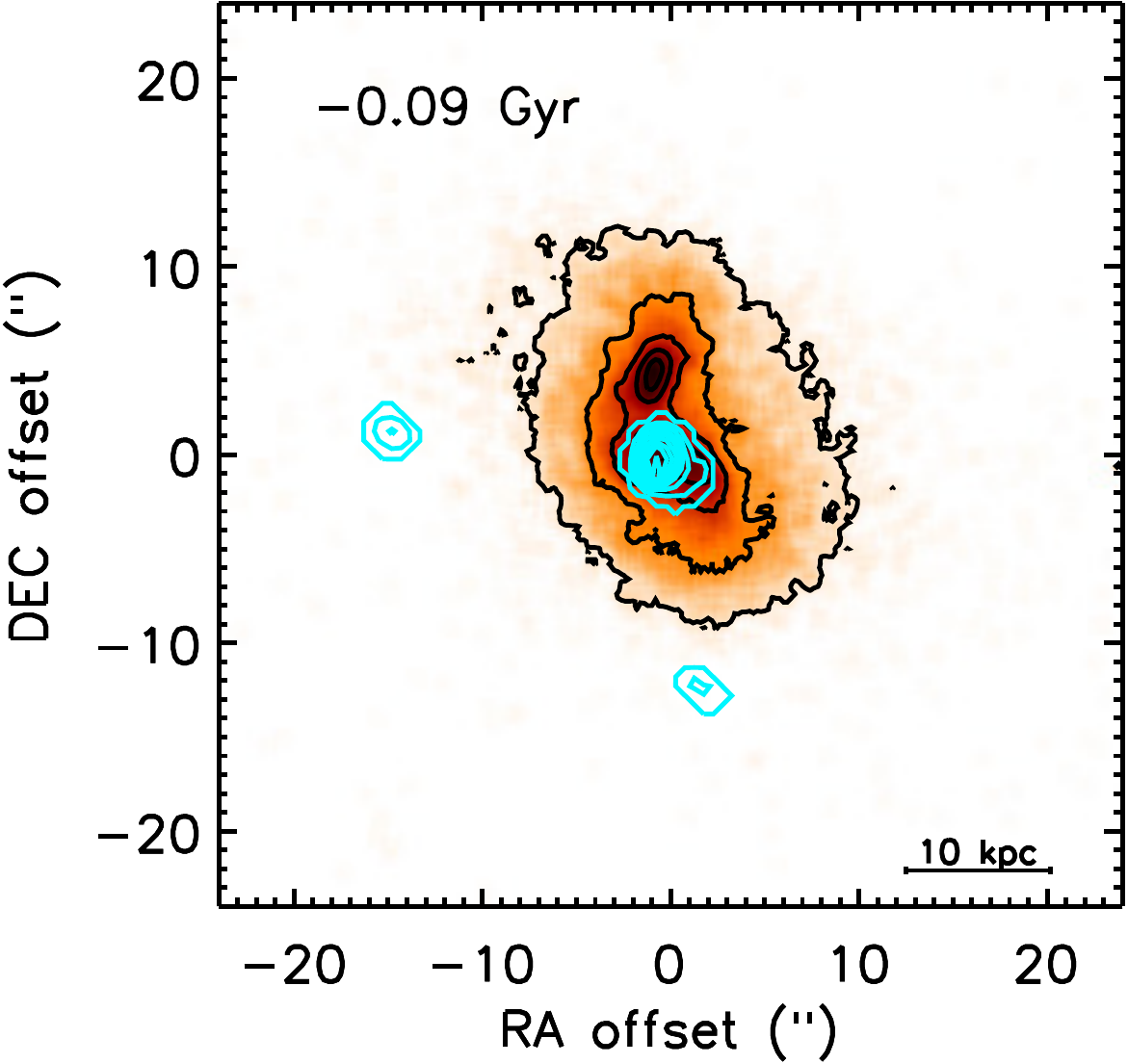}\includegraphics[height=4.85cm,angle=0,clip,trim=2.2cm 0cm 0cm 0cm]{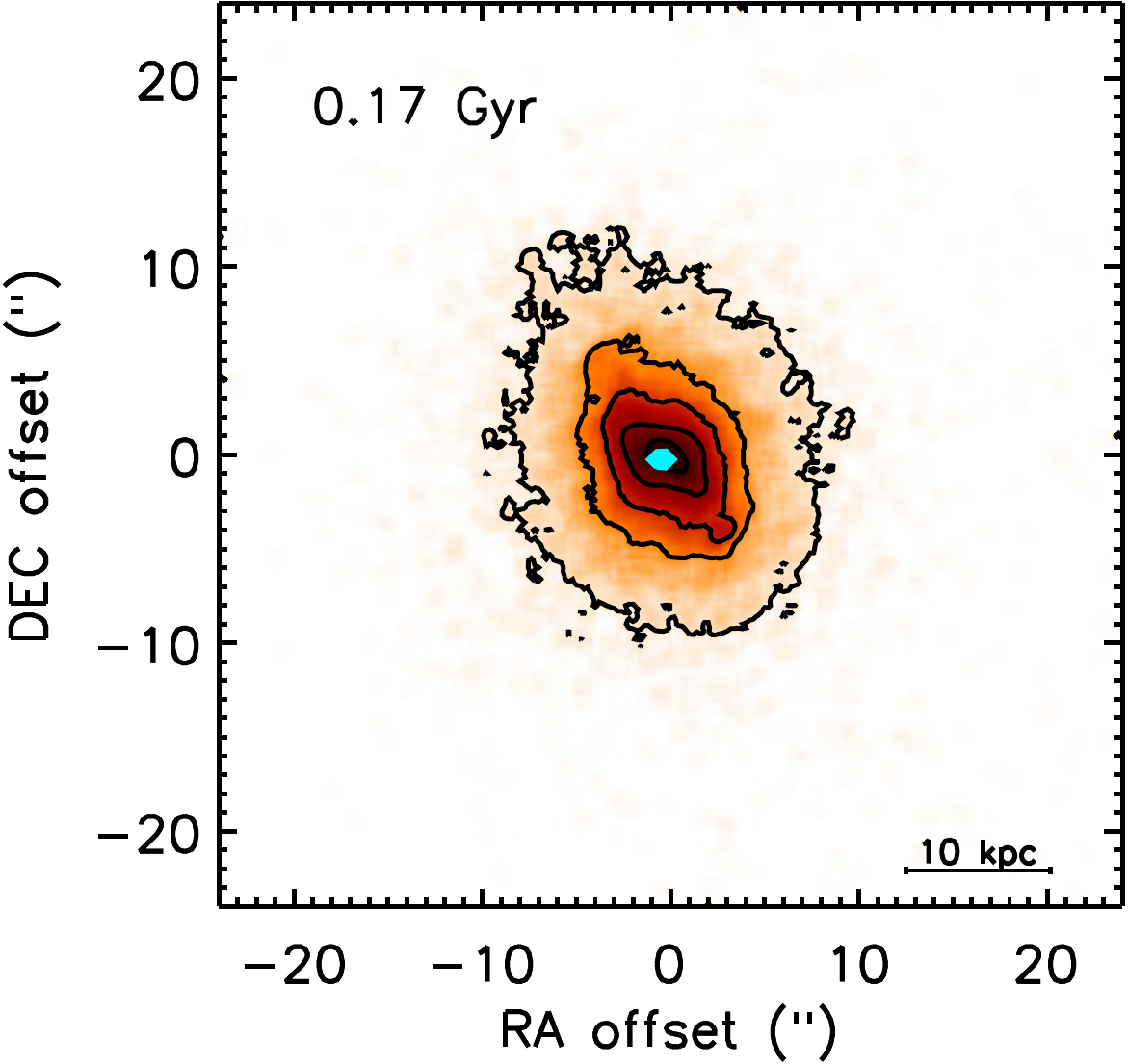}\includegraphics[height=4.85cm,angle=0,clip,trim=2.2cm 0cm 0cm 0cm]{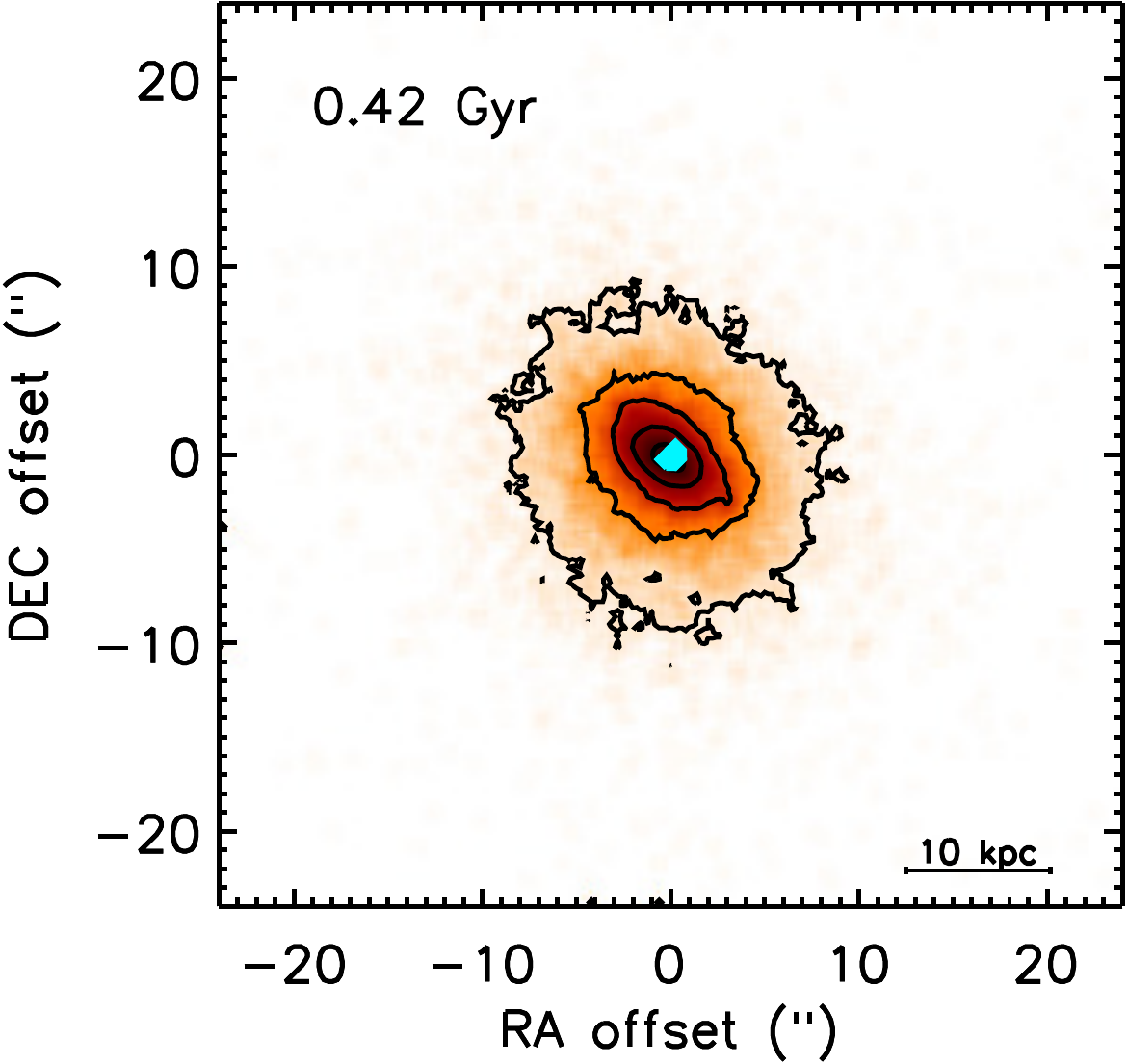}
 \caption{Mock observations of PSB385;  post-starburst episode induced by a merger.}
    \end{subfigure}
\caption{Mock $^{12}$CO($J$=1-0) observations of two simulated PSBs (blue contours), overlaid on the stellar distribution of the object (orange, black contours) at several epochs during their PSB episode (indicated in the legend of each panel, with times relative to $t_{\rm PSB}$). Each galaxy was artificially redshifted to $z=0.07$ (the mean redshift of our observational comparison sample) and mock observed at 2 kpc (1\farc5) resolution with parameters matching typical ALMA observing setups. The top object (PSB29) undergoes ram pressure stripping of its ISM as it enters a larger halo. The bottom object (PSB385) undergoes a merger which triggers a starburst. PSBs are typically disturbed in the buildup to the PSB epsiode, and have compact (but still kinematically disturbed) morphologies afterwards.}
\label{mock_obs}
 \end{center}
 \end{figure*}
 
\subsection{Mock observations}

Gas-rich PSB galaxies are prime targets for observations with ALMA and other millimetre interferometers, and hence it is interesting to consider what we expect to see in these systems. In Figure~\ref{mock_obs} we show a time series of mock $^{12}$CO($J$=1-0) observations of two of our simulated PSBs during their PSB episode. 

In order to create these we extracted the positions, velocities and masses of the star-forming gas particles from the EAGLE snipshots, and fed these as inputs to the \textsc{KINematic Molecular Simulation (KinMS)} routines of \cite{2013MNRAS.429..534D}. We artificially redshifted each object to $z=0.07$ (the mean redshift of our observational comparison sample), where 1\arcsec $\approx$1.3 kpc with our adopted cosmology. We performed mock observations using the \textsc{KinMS} tool with an observation setup typical of ALMA observations (a 1\farc5 beamsize, 0\farc5 pixels, 10 \kms\ channels and a sensitivity of $\approx$1.5 mJy/beam). 

We chose two objects that have PSB episodes at low-redshift, and follow two of the typical evolutionary pathways we discussed above. The object in the top panel of Figure \ref{mock_obs} (PSB29) undergoes ram pressure stripping of its ISM as it enters a larger halo. The object in the bottom panel (PSB385) undergoes a merger which triggers a starburst. This figure reflects the results discussed above: the gas in PSB is are typically disturbed in the buildup to the PSB episode, and has a compact (but still kinematically disturbed) morphology shortly thereafter.

Given the size evolution of PSBs (shown in Figure \ref{sfe_evol_size}) it would seem that $\approx$one kiloparsec or better resolution is likely required to truly resolve the molecular gas reservoirs of typical PSB galaxies at low redshifts. We would predict that the majority of PSB galaxies will have disturbed gas, which is slowly beginning to settle back into the potential of the host galaxy (see Section \ref{kappa_sec}).

\section{\uppercase{Discussion and Conclusions}}
\label{discuss}
\label{conclude}

\noindent In this paper we have selected PSB galaxies from the EAGLE (100 Mpc)$^3$ reference simulation, by searching for objects with a strong drop in their sSFR over a period of 600 Myr. These objects match the space density and stellar mass distribution of observed high-mass PSB galaxies, and have realistic sizes. After post-processing the simulation to create mock spectra we find that these objects would make it into spectroscopic PSB galaxy samples, if observed at the correct phase in their evolution. These successes give us confidence that we can use this sample of simulated galaxies to study the cold gas properties of PSB galaxies.

The vast majority of our simulated PSBs have significant gas reservoirs, with median star-forming gas masses of $\approx2\times10^{9}$\,\msun\ at low redshift. At all redshifts $<$5 per cent of PSBs are completely devoid of star-forming gas (at the mass resolution of our simulation). We find only a small amount of evolution in the median gas content of PSB galaxies with cosmic time, evolving as $\sim(1+z)$ rather than $\approx(1+z)^{2.5}$ for the evolution of the total molecular gas content of galaxies in this redshift range \citep[e.g.][]{2017ApJ...835..120M}. Once one includes selection and volume effects we find good agreement between the observed and simulated gas masses of the PSB population.

The power of studying simulated PSB galaxies is that one can examine the time evolution of their properties, and compare this to observations (which necessarily reveal the properties of  any one object only at a single time). This allows us to validate and interpret the observations, and guide future studies. We showed here that EAGLE well reproduces the observed time evolution of the gas fraction of the PSB galaxy population -- with the average galaxy losing $\approx$90 per cent of its star-forming gas in only $\approx$600 Myr.  The excellent reproduction of this observed trend in the simulations strengthens our confidence in this result, as well as lending additional credence to observational timescale determinations.

This fast evolution in the gas fraction of PSB galaxies  is accompanied by a clear decrease in the efficiency of star formation. While not as dramatic as the observed star formation suppression seen in some PSBs (due to the mass limit and simple ISM model in EAGLE), the ubiquity of this lower SFE is striking. The processes which cause the PSB episode  initially lead to a compaction of the gas disc, as a starburst occurs (or sometimes, at low redshift, as environmental effects remove the outer gas reservoir). As the PSB galaxy evolves the reservoir expands somewhat, leading to a lower dense gas fraction. Although a direct comparison is difficult, the sense of this correlation matches the observations of low HCN-to-CO ratios in PSB galaxies \cite{2018arXiv180512132F}. 

We showed that the mechanisms causing galaxies to become PSBs are quite diverse, as are the mechanisms which quickly deplete the gas reservoirs of these systems. 
\begin{itemize}
\item Both major (1:1 to 1:3) and minor mergers ($<$1:3) are important in creating PSBs. Major mergers are implicated most often at low redshift, while the more common minor mergers dominate at higher redshift. These mergers can create starbursts, and trigger AGN activity, both of which deplete the gas reservoir very quickly. Mergers can also throw tidal tails of gas to large radii, which at a later time can fall back and refuel the resultant galaxy. 
\item  Environmental effects are also important, with $\approx$20 per cent of PSBs becoming a satellite around the time they exhibit PSB features, and with at least 50 per cent of these systems clearly suffering the effects of ram-pressure stripping. 
\item Some AGN activity is present in almost all of the simulated PSB galaxies. However, we showed that this activity is somewhat sub-dominant, with less than 10 per cent of simulated PSBs having a strong AGN outburst during the PSB period. These outbursts are likely triggered by other mechanism(s) which have already created the PSB. Thus although it can help with the removal of gas, AGN activity does not seem to be an important way to create PSB galaxies in EAGLE. AGN also cannot be dominantly responsible for the depletion of the gas reservoirs in PSB galaxies, as high accretion rates are not present in the majority of PSB galaxies. 
\end{itemize}

Finally we predict that, once high-resolution resolved images of the molecular gas in PSBs are available, they will show that the vast majority of PSB systems have disturbed gas kinematics. Within the simulation the fraction of PSBs that show dominant disc rotation ($\kappa_{\rm gas}>0.8$) is very low ($<5$ per cent). We also showed that fairly compact gas reservoirs should be expected, especially in recently quenched PSB galaxies. These small sizes persist for a long period of time at low redshifts, consistent with the small (physical) sizes of the residual cold gas reservoirs in quenched galaxies \citep[e.g.][]{2013MNRAS.429..534D}. 

We conclude that the evolution of PSB galaxies within the EAGLE (100 Mpc)$^3$ reference simulation is fully consistent with recent observations of molecular gas rich PSB galaxies. The mechanisms causing galaxies to become PSBs are diverse, but lead to a strikingly similar post-burst evolution, with a fast loss of gas and a low star formation efficiency that lasts for at least a gigayear. Further observational and theoretical work on such systems is crucial in order to reveal in detail the balance between different gas depletion mechanisms, and the physical processes suppressing the efficiency of star formation in such systems as they evolve towards the red sequence. 

\vspace{0.5cm}
\noindent \textbf{Acknowledgments}

We would like to thank Decker French, Ben Forrest, Thomas Guillet and R\"udiger Pakmor for helpful discussions and/or for sharing their data. We also thank all involved with the conference entitled ``The Physics of Quenching Massive Galaxies at High Redshift'', held 6 Nov 2017 through 10 Nov 2017 at the Lorentz Centre, Leiden, which inspired this study. 
TAD acknowledges support from a Science and Technology Facilities Council Ernest Rutherford Fellowship. Support for FvdV was provided
by the Klaus Tschira Foundation.
The work of SM was supported by the Science and Technology Facilities Council (grant number ST/P000541/1) and the Academy of Finland, grant number: 314238.
RAC is a Royal Society University Research Fellow.

This work used the DiRAC Data Centric system at Durham University, 
operated by the Institute for Computational Cosmology on behalf of the 
STFC DiRAC HPC Facility. This equipment was funded 
by BIS National E-infrastructure capital grant ST/K00042X/1, STFC capital 
grant ST/H008519/1, and STFC DiRAC Operations grant ST/K003267/1 and 
Durham University. DiRAC is part of the National E-Infrastructure (\url{www.dirac.ac.uk}).

\bsp
\bibliographystyle{mnras}
\bibliography{bib_psb}
\bibdata{bib_psb.bib}
\bibstyle{mnras}

\label{lastpage}
\end{document}